\documentclass[12pt,a4paper,final]{iopart}

\usepackage[titletoc,title]{appendix}
\usepackage{iopams}  
\usepackage{graphicx}
\usepackage{xcolor}
\expandafter\let\csname equation*\endcsname\relax
\expandafter\let\csname endequation*\endcsname\relax
\usepackage{amsmath}
\usepackage[breaklinks=true,colorlinks=true,linkcolor=blue,urlcolor=blue,citecolor=blue]{hyperref}

\begin{document}
\title[Advanced LIGO - Fermi-GBM Single Interferometer Search]{Search for Advanced LIGO Single Interferometer Compact Binary Coalescence Signals in Coincidence with Gamma-Ray Events in Fermi-GBM}

\author{C Stachie$^1$, T Dal Canton$^{2,3,4}$, E Burns$^2$, N Christensen$^{1,5}$, R Hamburg$^{6,7}$, M Briggs$^8$, J Broida$^5$, A Goldstein$^8$, F Hayes$^9$, T Littenberg$^8$, P Shawhan$^{10}$, J Veitch$^9$, P Veres$^8$, C A Wilson-Hodge$^{11}$}
\address{$^1$ Artemis, Universit\'e C\^ote d'Azur, Observatoire C\^ote d'Azur, CNRS, CS 34229, F-06304 Nice Cedex 4, France}
\address{$^2$ NASA Goddard Space Flight Center, Greenbelt, MD 20771, USA}
\address{$^3$ Max-Planck-Institut f\"ur Gravitationsphysik (Albert Einstein Institut), Am M\"uhlenberg 1, D-14476 Potsdam-Golm, Germany}
\address{$^4$ Universit\'e Paris-Saclay, CNRS/IN2P3, IJCLab, 91405 Orsay, France}
\address{$^5$ Physics and Astronomy, Carleton College, Northfield, MN 55057, USA}
\address{$^6$ Department of Space Science, University of Alabama in Huntsville, Huntsville, AL 35899, USA}
\address{$^7$ Center for Space Plasma and Aeronomic Research, University of Alabama in Huntsville, Huntsville, AL 35899, USA}
\address{$^8$ Science and Technology Institute, Universities Space Research Association, Huntsville, AL 35805, USA}
\address{$^9$ Institute  for  Gravitational  Research,  University  of  Glasgow,  Glasgow,  G12  8QQ, UK}
\address{$^{10}$ Department of Physics, University of Maryland, College Park, MD 20742, USA}
\address{$^{11}$ NASA Marshall Space Flight Center, Huntsville, AL 35812, USA}
\ead{cosmin.stachie@oca.eu}

\begin{abstract}
\setlength{\parindent}{0cm}

Presented is the description of a new and general method used to search for $\gamma$-ray counterparts to gravitational-wave (GW) triggers. This method is specifically applied to single GW detector triggers. Advanced LIGO data from observing runs O1 and O2 were analyzed, thus each GW trigger comes from either the  LIGO-Livingston or the LIGO-Hanford interferometer. For each GW trigger,  {\it Fermi} Gamma-ray Burst Monitor data is searched and the most significant subthreshold signal counterpart is selected. Then, a methodology is defined in order to establish which of the GW-$\gamma$-ray trigger pairs are likely to have a common origin. For that purpose an association ranking statistic is calculated from which a false alarm rate is derived. The events with the highest ranking statistics are selected for further analysis consisting of LIGO detector characterization and parameter estimation. The $\gamma$-ray signal characteristics are also evaluated. We find no significant candidates from the search.
\end{abstract}

\pacs{00.00, 20.00, 42.10}
\vspace{2pc}
\noindent{\it Keywords}: LIGO, GBM, Fermi, PyCBC

\submitto{\CQG}

\section{Introduction}

Advanced LIGO~\cite{TheLIGOScientific:2014jea} and Advanced Virgo~\cite{TheVirgo:2014hva} are km-scale interferometers dedicated to the detection of gravitational waves (GWs). Since the start of the advanced detector era in 2015, several compact binary coalescence (CBC) events have been detected~\cite{2018arXiv181112907T}. The detection of GW170817~\cite{PhysRevLett.119.161101}, a binary neutron star merger in coincidence with electromagnetic (EM) waves, enabled a huge step forward in understanding these cataclysmic events~\cite{abbott2017gravitational,GBM:2017lvd,Goldstein_2017b, Savchenko_2017}. While GWs encode information related to the dynamics of the binary system and to the characteristics of the compact objects, like masses and spins, EM radiation gives precious insight into the behaviour of matter in extreme environments. $\gamma$-ray information in particular is linked directly to the local environment. The detection of both the GW and EM signals originating from a compact binary merger allows one to address questions related to fundamental physics, like the speed of gravity calculation~\cite{abbott2017gravitational}, a measurement of the nuclear equation of state~\cite{Abbott:2018exr}, and constraining  the Hubble constant~\cite{Abbott:2017xzu}.

GW CBC ``triggers'' identified in an interferometer's data are characterized by a matched-filter signal-to-noise ratio (SNR) which would be the optimal detection statistic in stationary Gaussian noise. However, one of the big challenges in LIGO-Virgo data analysis is to distinguish non-Gaussian and non-stationary noise transients from astrophysical transients. For a given CBC trigger, its false alarm rate (FAR), representing how often a noise event like this or more significant (meaning by measurement of FAR) is detected, provides a means to address this obstacle, but the calculation of a FAR relies either on time-shifting two or more detector data streams or on modeling the noise properties. However a simultaneous detection between a single interferometer GW signal and some multimessenger counterpart, for instance an EM or neutrino event, could increase the statistical confidence of the GW signal.  It is worth mentioning that although we are dealing with single interferometer LIGO triggers and Fermi-GBM candidates in this study, this approach is general and can be applied in various other cases.

The analysis method presented in this paper is intended to be generalizable to any two types of multimessenger events, provided that each signal comes out with its own statistical significance and some correlation is expected between the two signals, such as the same time of arrival and/or the same spatial origin. Thereby one could consider associations between two of the following different astrophysical signals: triggers from a GW search pipeline, $\gamma$-ray burst (GRB) prompt emission or high energy neutrinos. Although there is a high degree of generality for the method presented in this paper, the study here is focused on the case of joint detections between PyCBC~\cite{Usman_2016,Nitz:2017} single interferometer GW triggers and Fermi-GBM $\gamma$-ray signals.

The Fermi Gamma-ray Space Telescope~\cite{Meegan_2009,Michelson:2010zz} is a space observatory dedicated to the detection of the most energetic phenomena taking place in the universe through observations of $\gamma$-ray radiation. Aboard Fermi, the Gamma-ray Burst Monitor (GBM) instrument~\cite{Meegan_2009} is used to observe GRBs. GRBs are traditionally classified in two categories: long GRBs~\cite{Woosley:2006fn} which are supposed to be associated with a sub-class of core-collapse supernovae, and short GRBs~\cite{article_Nakar} which are believed to originate in CBC systems. While the search for EM counterparts to binary neutron star (BNS) and neutron star--black hole (NS-BH) mergers is motivated by both theoretical studies and experimental observations, the GW150914-GBM event, possibly associated with a binary black hole (BBH) merger~\cite{Connaughton_2016, Greiner_2016, Connaughton_2018}, provides a motivation to also follow-up BBH GW signals for EM counterparts.

In the last few years, several GW search pipelines were designed in order to target CBC signals buried in the GW interferometer data. To this end, two kinds of pipelines were developed: modelled searches~\cite{Usman_2016,Sachdev:2019vvd,2016CQGra..33q5012A, Hooper:2011rb} which look specifically for signals from CBCs, and unmodeled (burst) searches~\cite{PhysRevD.93.042004} that aim to detect a broader range of astrophysical phenomena such as core-collapse of massive stars, magnetar star-quakes, and CBCs. For the present study, we limit the analysis to GW triggers provided by the PyCBC pipeline~\cite{Usman_2016,Nitz:2017}. PyCBC is a modeled pipeline which identifies CBC signals by performing a matched-filter search using a bank of GW template waveforms~\cite{Usman_2016,Nitz:2017}. The Fermi-GBM follow-up is realized using a tool called the GBM Targeted Search \cite{Blackburn:2014rqa,Targeted_Search_Kocevski_2018}. The Targeted Search version used for this study is from~\cite{Goldstein:2016zfh}.

While recently a search method for Fermi-GBM counterparts to LIGO single interferometer BNS candidates was presented~\cite{Nitz_2019}, the present study introduces a follow-up of all single-detector CBC candidates, regardless of the properties of the originating compact objects. We focus here on the analysis of Advanced LIGO data from the O1 and O2 observing runs with GW triggers produced by the PyCBC pipeline, although the method can be generalized. In addition, this paper serves as a technical accompaniment to the comprehensive search for coincident GW and $\gamma$-ray triggers during O1 and O2 from LIGO-Virgo and Fermi GBM~\cite{Hamburg:2020dtg}.

This paper is structured as follows: we start with a brief description of the LIGO and Fermi-GBM triggers in Section 2. In Section 3 we show our derivation of the joint ranking statistic $\Lambda$. 
A procedure to get a FAR distribution with respect to $\Lambda$ is presented in Section 4. Section 5 summarizes the results of this search using O1 and O2 data, and we conclude this study in Section 6. Appendix A motivates some parameters chosen in Section 2, and Appendix B describes the impact of a fitting procedure on the final results. 

\section{LIGO and Fermi-GBM triggers}

We begin our search with a set of input single-detector GW triggers from the LIGO-Hanford and LIGO-Livingston detectors. We take the triggers from the PyCBC analysis given in the GWTC-1 catalog~\cite{2018arXiv181112907T}, which covers the search space described in~\cite{DalCanton:2017ala} and hence include potential BNS, NSBH and BBH signals. Each trigger is ranked by a statistic $\hat{\rho}_{gw}$, a combination of the trigger's matched-filter signal-to-noise ratio and two $\chi^2$ signal-based vetoes~\cite{Usman_2016, Allen:2004gu}. We keep only those triggers having $\hat{\rho}_{gw} \geq 8$. More details about this choice are given in \ref{sec:lower_limits}.

For each GW trigger, we analyze nearby Fermi-GBM time-tagged event data using the Targeted Search~\cite{Blackburn:2014rqa,Targeted_Search_Kocevski_2018}. The Targeted Search looks for excesses of photon counts compatible with GRBs over a variety of overlapping time windows $\pm 30$ s from the input GW trigger time, using search timescales from 0.256 s to 8.192 s. For each time window, a log-likelihood ratio (LLR) is computed. The LLR accounts for the fact that the photon rates produced by a GRB in the GBM detectors and energy channels are not independent, but can be predicted after a particular spectral shape has been assumed for the GRB. We generate GBM ``triggers'' by only keeping the window having the highest LLR if it fulfills the condition LLR $\geq 5$. The choice of this lower limit is motivated in \ref{sec:lower_limits}.

The next tasks are to identify pairs of GW-GBM triggers which could plausibly originate from a common astrophysical event, find a way to rank the pairs, and assign a statistical significance to them.

\section{Association ranking statistic}
\label{sec:association ranking}

The main ideas and techniques used here are an extension of the Bayesian formalism introduced in \cite{Ashton_2018}. 
We note by $D_{L}$ and $D_{G}$ the data sets from LIGO and Fermi-GBM, respectively, and consider the following hypotheses: ($H^{C}$) both data sets contain a transient signal and the two signals are emitted by a common source; ($H^{NN}$) both data sets contain only noise; ($H^{SN}$) there is a signal in LIGO data and only noise in Fermi-GBM data; ($H^{NS}$) there is only noise in LIGO data and a signal in Fermi-GBM data; and ($H^{SS}$) both data sets contain signals, but the signals come from unrelated sources. The joint ranking statistic considered hereafter is the Bayes factor comparing the astrophysically interesting hypothesis $H^{C}$ against the logical disjunction of all other hypotheses:
\begin{equation}
    \Lambda = \frac{P(D_L, D_G | H^C)}{P(D_L, D_G | H^{NN} \vee H^{SN} \vee H^{NS} \vee H^{SS})}.
\end{equation}
This expression can be factorized as
\begin{eqnarray}
\Lambda & = & \frac{P(D_L, D_G | H^C)}{P(D_L, D_G | H^{NN} \vee H^{SN} \vee H^{NS} \vee H^{SS})} \\
        & = & \frac{P(D_L, D_G | H^C)}{P( H^{NN} \vee H^{SN} \vee H^{NS} \vee H^{SS} | D_L, D_G) \cdot \frac{P(D_L, D_G)}{P( H^{NN} \vee H^{SN} \vee H^{NS} \vee H^{SS})}} \\
        & = & \frac{P(D_L, D_G | H^C) \cdot \sum_{\substack{X, Y \in \{N, S\}}}P(H^{XY})}{\sum_{\substack{X, Y \in \{N, S\}}} P(H^{XY} | D_L, D_G) \cdot {P(D_L, D_G)}} \\
        & = & \frac{P(D_L, D_G | H^C) \cdot \sum_{\substack{X, Y \in \{N, S\}}}P(H^{XY})}{\sum_{\substack{X, Y \in \{N, S\}}} P( D_L, D_G | H^{XY}) \cdot {P(H^{XY})}} \\
        & = & \frac{4 \cdot P(D_L, D_G | H^C)}{\sum_{\substack{X, Y \in \{N, S\}}} P( D_L, D_G | H^{XY})} \\
        & = & \frac{4}{\sum_{\substack{X, Y \in \{N, S\}}} \frac{1}{\mathcal{B}_{C/XY}(D_L, D_G)}},
\end{eqnarray} 
where by $\mathcal{B}_{C/XY}(D_L, D_G) = P(D_L, D_G | H^C) / P(D_L, D_G | H^{XY})$  we note the likelihood ratio of the hypothesis $H^C$ and $H^{XY}$.
Equations (2) and (4) are obtained by means of Bayes theorem and the derivation of Equation (5) needs the equal priors assumption $P(H^C) = P(H^{XY})$ $\forall X, Y \in \{N, S\}$. Although at first glance the equal prior assumption can appear unrealistic, it can be justified as follows. On the one hand, it is the choice that makes the calculation simplest. On the other hand, because we will eventually convert $\Lambda$ to a frequentist FAR (described in Section 4), its strict interpretation as a Bayes factor is relatively unimportant. Following the same procedure as \cite{Ashton_2018}, in particular using the assumption $P(D_{L/G}|H^{c}) = P(D_{L/G}|H^s)$, one has
\begin{eqnarray}
    \mathcal{B}_{C/NN} &=& \frac{I_{\Delta t} I_{\mathbf{\Omega}}} {Q_L Q_G} \\ \mathcal{B}_{C/SN} &=& \frac{I_{\Delta t} I_{\mathbf{\Omega}}} {Q_G} \\ \mathcal{B}_{C/NS} &=& \frac{I_{\Delta t} I_{\mathbf{\Omega}}} {Q_L} \\ \mathcal{B}_{C/SS} &=& I_{\Delta t} I_{\mathbf{\Omega}}
\end{eqnarray}
where $Q_L = Q_L(D_L) = P(D_L | \textrm{noise}) / P(D_L | \textrm{signal})$ and $Q_G = Q_G(D_G) = P(D_G | \textrm{noise}) / P(D_G | \textrm{signal})$ are the single-instrument Bayes factors comparing the noise-only and noise-plus-signal hypotheses in LIGO and GBM, respectively. $I_{\Delta t}$ and $I_{\mathbf{\Omega}}$ quantify the overlap of the posterior distributions for the arrival times (time offset) and sky locations (skymap overlap) inferred separately from the GW and $\gamma$-ray data.
Finally, by ignoring the overall factor of 4, the expression of joint ranking statistic becomes 
\begin{equation}
    \Lambda = \frac{I_{\Delta t} I_{\mathbf{\Omega}}}{1 + Q_L + Q_G +Q_L Q_G}.
    \label{eq:jointrank}
\end{equation}
We are allowed to drop the 4 factor because the numerical value of $\Lambda$ does not need to have a firm statistical meaning given that we ultimately form a background distribution of $\Lambda$ and use that to empirically assign a FAR. That is to say, we can consider any expression for $\Lambda$ as long as we do the same for the background and foreground.

 In order to evaluate $\Lambda$ for a specific pair of LIGO and Fermi-GBM triggers, one needs to calculate its four defining quantities from the properties of the triggers.
 Before showing how one can handle the computation of these different quantities, we emphasize some intuitive behavior of the joint ranking statistic (\ref{eq:jointrank}). The noise against signal Bayes factors $Q_L$ and $Q_G$ are decreasing functions with respect to the statistical significance of the individual LIGO and Fermi-GBM candidates. If both candidates have low significance (large $Q$), then $\Lambda \propto I_{\Delta t} I_{\mathbf{\Omega}} / (Q_L Q_G)$, which is small. If only one candidate of the pair, say the LIGO trigger, has very high statistical significance, then $Q_L \ll 1$ and $\Lambda \propto  I_{\Delta t} I_{\mathbf{\Omega}} / Q_G$, i.e. the joint ranking statistic depends in some sense only on the significance of the other candidate and on the time and skymap overlap. Finally if both candidates are very statistically significant then $\Lambda \propto I_{\Delta t} I_{\mathbf{\Omega}}$, i.e.~the compatibility of the arrival times and sky locations becomes the only relevant metric.

\begin{figure}[!htb]
   \begin{minipage}{0.55\textwidth}
     \centering
     \includegraphics[width=\linewidth]{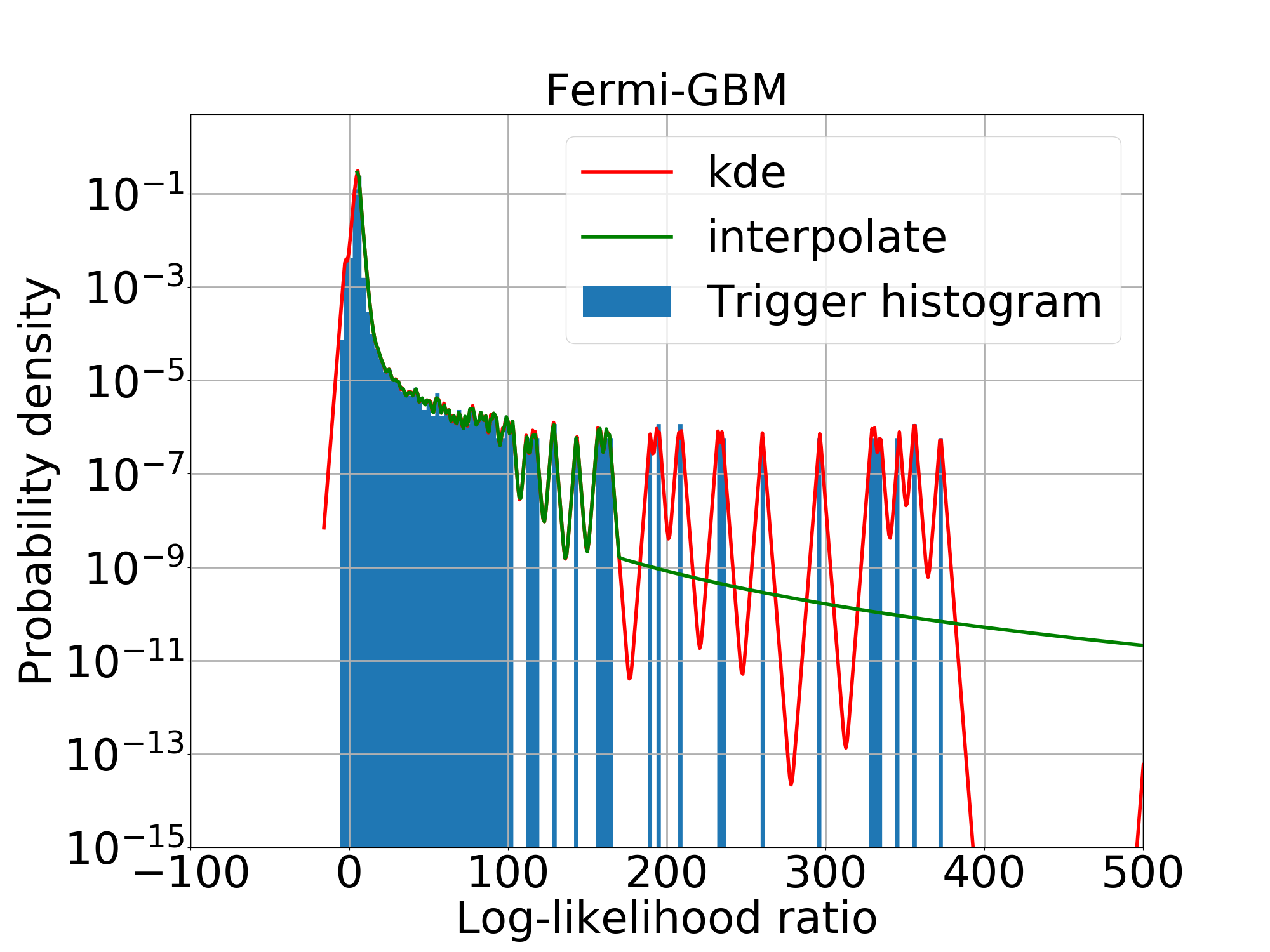}
   \end{minipage}\hfill
   \begin{minipage}{0.55\textwidth}
     \centering
     \includegraphics[width=\linewidth]{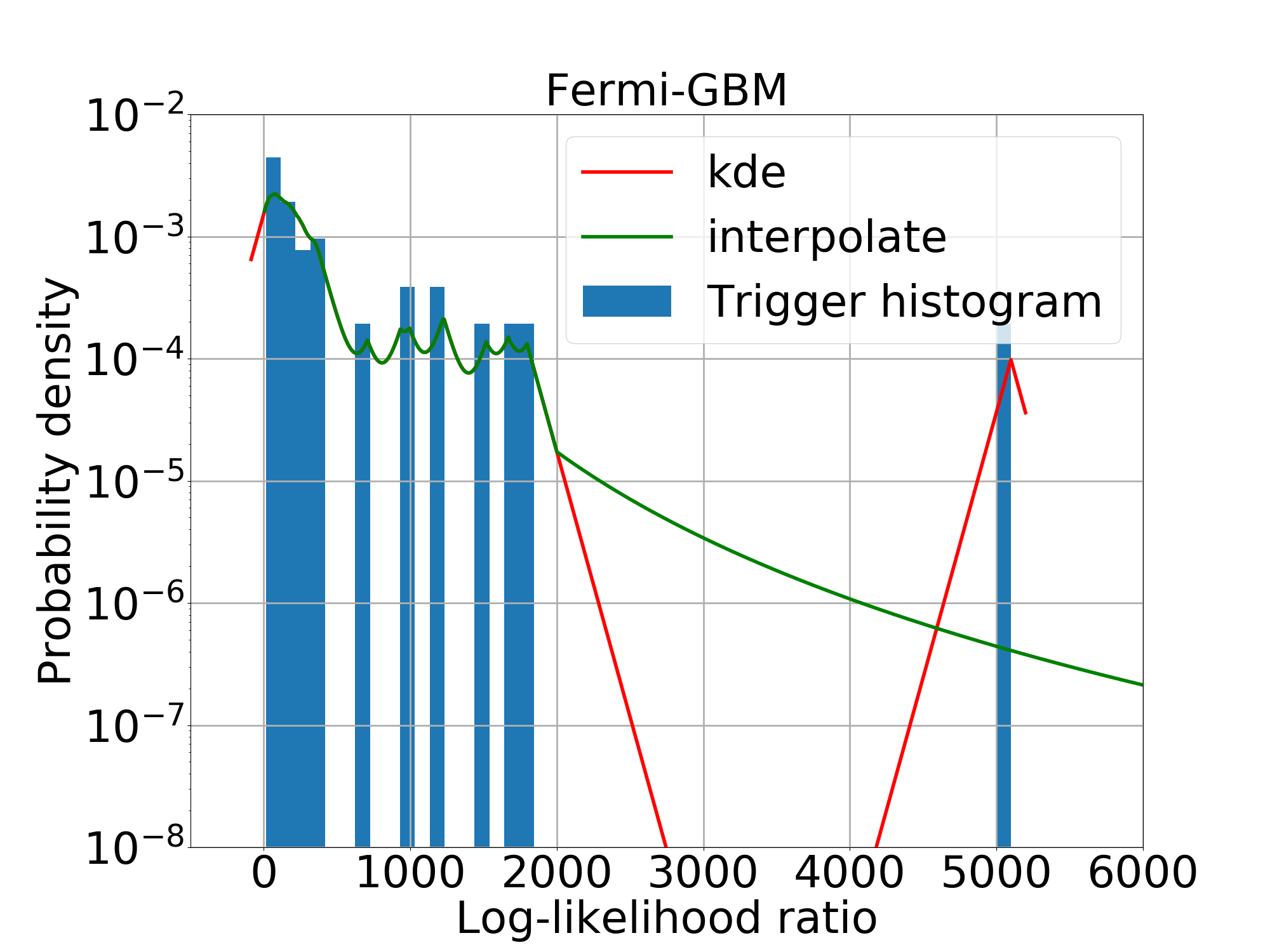}
   \end{minipage}
   \caption{The steps realized to generate $P(LLR|\textrm{noise})$ (at left) and $P(LLR|\textrm{signal})$ (at right). The histogram of triggers with respect to the LLR is illustrated on solid blue. The fitting using the KDE method is represented in red. A minimum and a maximum threshold are chosen to delimit the LLR range on which the KDE fitting is considered. Finally, the fitted curve is interpolated (on green) for the region in between the thresholds and a prior $\propto LLR^{-4}$ is chosen for high LLRs.}
   \label{Fig:kde gbm}
\end{figure}

In this study, we take the Fermi-GBM Bayes factor $Q_G$ to be a function uniquely dependent on the log likelihood ratio (LLR). This quantity compares the signal presence hypothesis against the null hypothesis of only background noise~\cite{Goldstein:2016zfh}. The dependence of $Q_G$ on LLR is given by $Q_G(LLR) = \frac{P(LLR|\textrm{noise})}{P(LLR|\textrm{signal})}$. As such, in order to get $Q_G(LLR)$, one needs the distribution of LLR with respect to noise and signals. A sample of real signals~\cite{Targeted_Search_Kocevski_2018} was used to create a histogram of $LLR$. The distribution was fit using a kernel density estimation (KDE) from $LLR=5$ (sufficiently small threshold in order to be sure of not missing any interesting event) to $LLR=2000$ (this threshold is imposed by the quality of the KDE fitting).  For higher $LLR$ we assume $P(LLR|\textrm{signal})\propto LLR^{-4}$. The choice of the prior is consistent with a uniformly distributed population of binaries in the universe and a $LLR$ inversely proportional to the distance, a fact supported by~\cite{Targeted_Search_Kocevski_2018}. For the distribution of noise $P(LLR|\textrm{noise})$, a histogram of Fermi-GBM backgrounds has been acquired during O2. Like in the case of signals, the histogram was fitted using KDE for values of LLR lower than a 170, then the prior $P(LLR|\textrm{noise})\propto LLR^{-4}$ was used for higher values of LLR. This time the choice of the $-4$ exponent is motivated by the wish of being conservative with what we have done for signals. The subsequent steps are illustrated in the Figure~\ref{Fig:kde gbm}.

\begin{figure}[!htb]
   \begin{minipage}{0.55\textwidth}
     \centering
     \includegraphics[width=\linewidth]{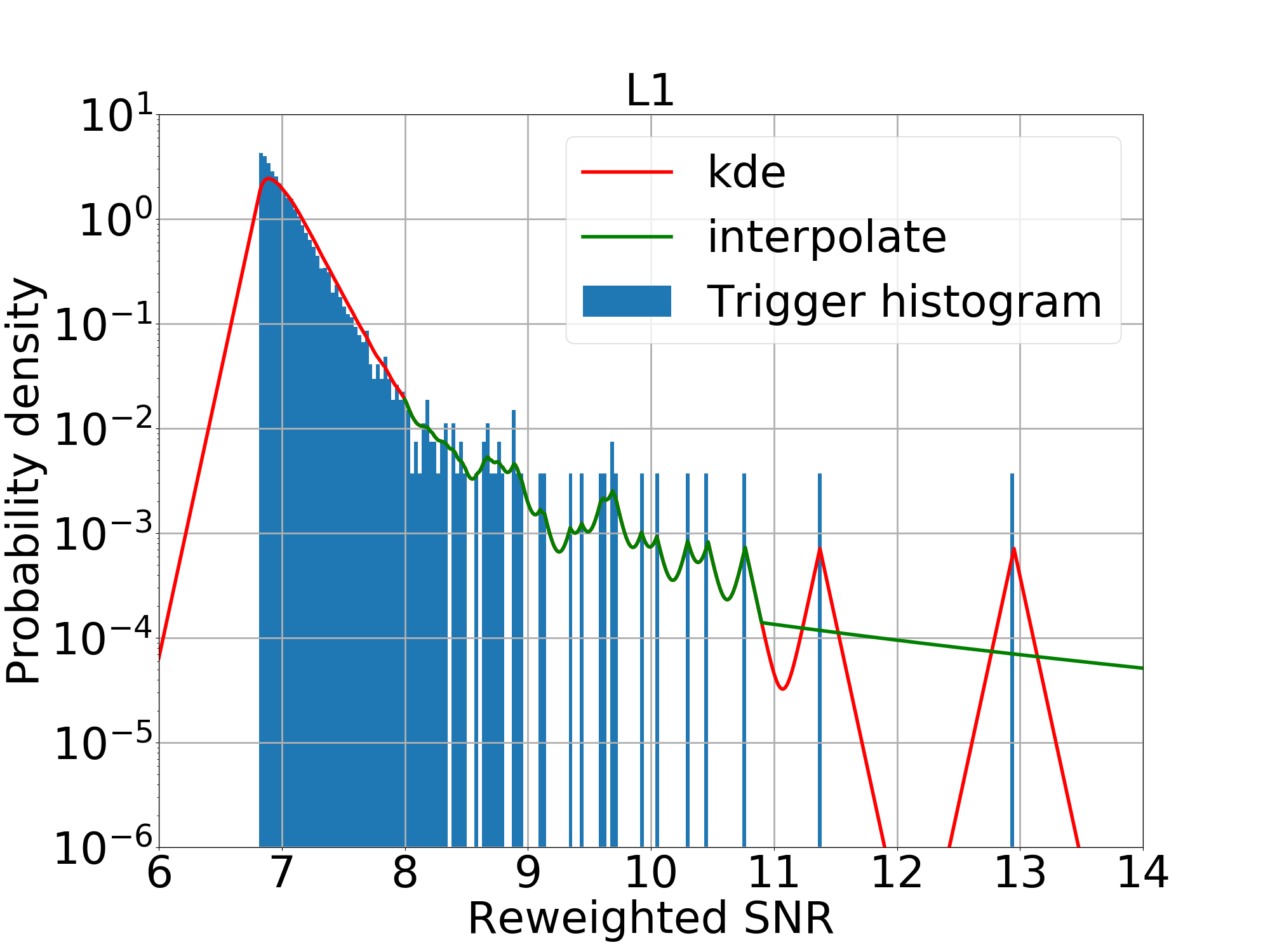}
   \end{minipage}\hfill
   \begin{minipage}{0.55\textwidth}
     \centering
     \includegraphics[width=\linewidth]{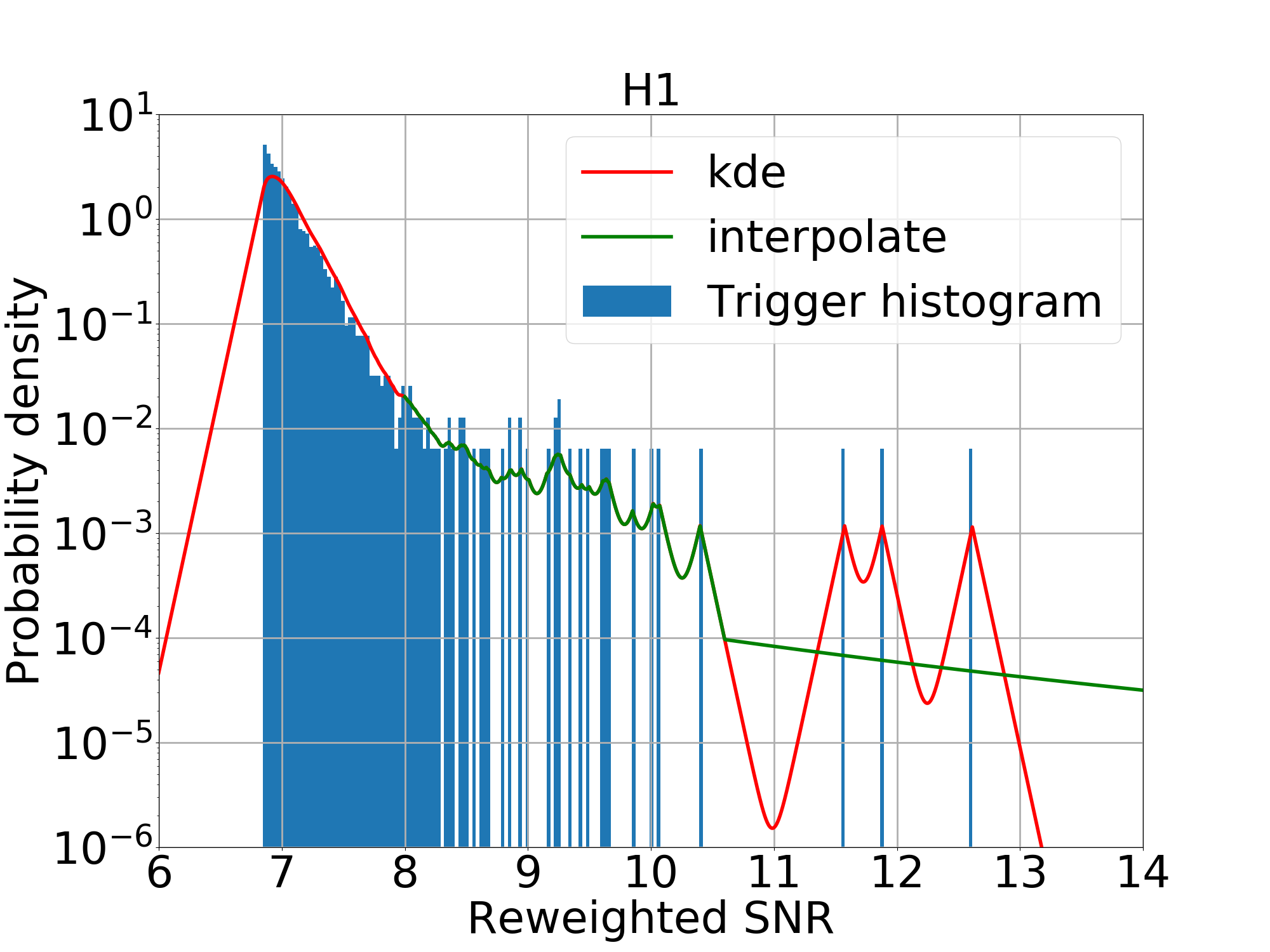}
   \end{minipage}
   \caption{Generation of $P(\hat\rho_{gw}|\textrm{noise})$ for L1 (at left) and H1 (at right) in O2. The different steps are illustrated: histogram of noise triggers (solid blue),  fitting of the underlying data (red), choice of thresholds and interpolation (green).}
   \label{Fig:stages}
\end{figure}

Concerning the LIGO Bayes factor, we choose the quantity to uniquely depend on $\hat\rho_{gw}$, a reweighted SNR which combines the matched-filter SNR with the $\chi^2$ veto~\cite{Usman_2016, Allen:2004gu} and with the high frequency sine-Gaussian $\chi^2$ discriminator presented in~\cite{Nitz:2017lco}. Therefore the expression for the LIGO Bayes factor is $Q_L(\hat\rho_{gw}) = \frac{P(\hat\rho_{gw}|\textrm{noise})}{P(\hat\rho_{gw}|\textrm{signal})}$. One needs the distributions of noise and signals for each interferometer. Again we start with a histogram of backgrounds, and then the histogram is fit. We introduce a minimum $\hat\rho_{gw}=8$ and a high threshold of $\hat\rho_{gw}=10.6$, and then we assume the prior $P(\hat\rho_{gw}|\textrm{noise}) \propto \hat\rho_{gw}^{-4}$ for higher $\hat\rho_{gw}$~\cite{Callister:2017urp}. As GW detections from only one interferometer have not been presented by LIGO and Virgo for observing runs O1 and O2~\cite{2018arXiv181112907T}, for the entire range of $\hat\rho_{gw}$ we assume $P(\hat\rho_{gw}|\textrm{signal}) \propto \hat\rho_{gw}^{-4}$. This process is done for each interferometer, LIGO-Livingston (L1) and LIGO-Hanford (H1), and for each observing run, O1 and O2. Figure~\ref{Fig:stages} shows the different stages in the generation of $P(\hat\rho_{gw}|\textrm{noise})$ during the observing run O2.

\begin{figure}[!htb]
   \begin{minipage}{0.55\textwidth}
     \centering
     \includegraphics[width=\linewidth]{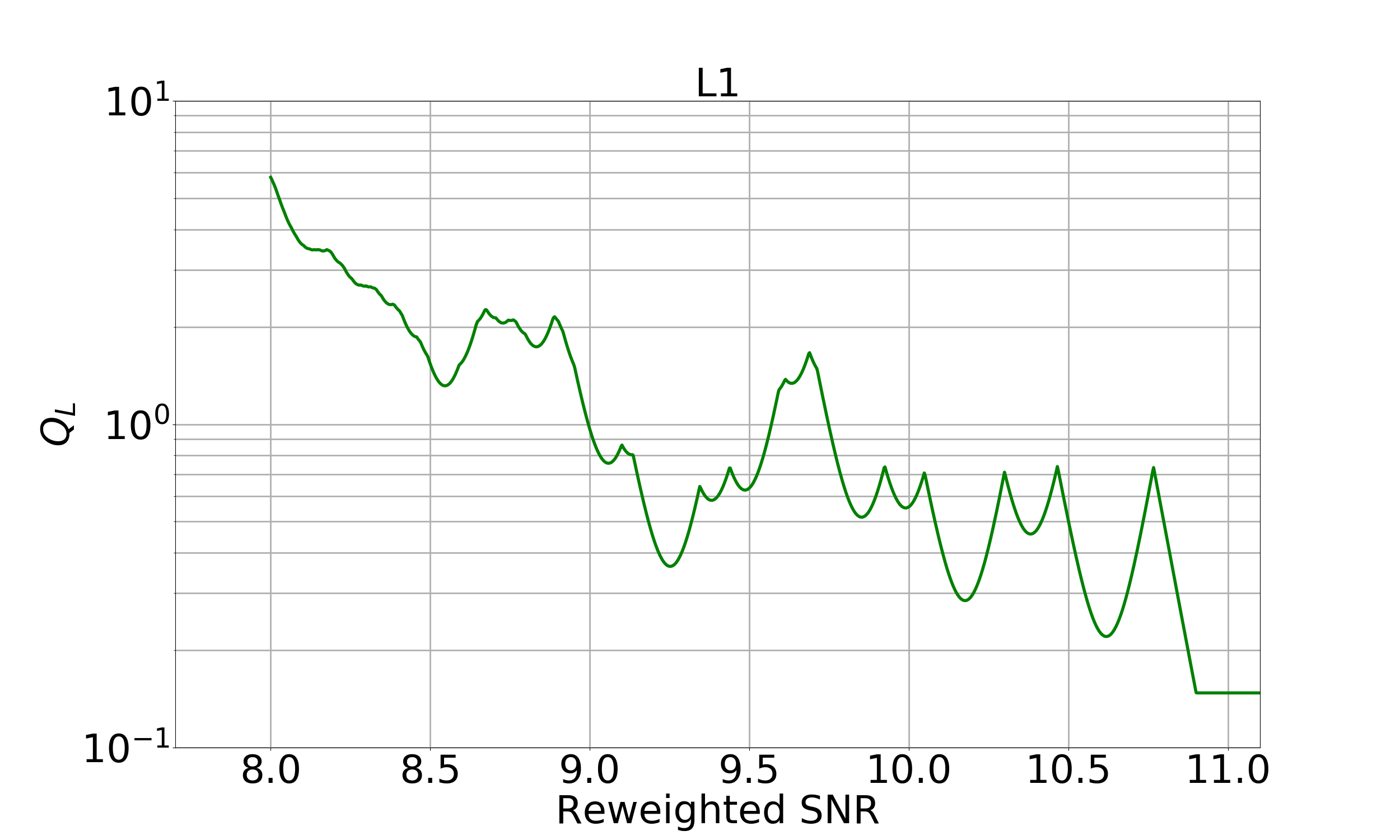}
   \end{minipage}\hfill
   \begin{minipage}{0.55\textwidth}
     \centering
     \includegraphics[width=\linewidth]{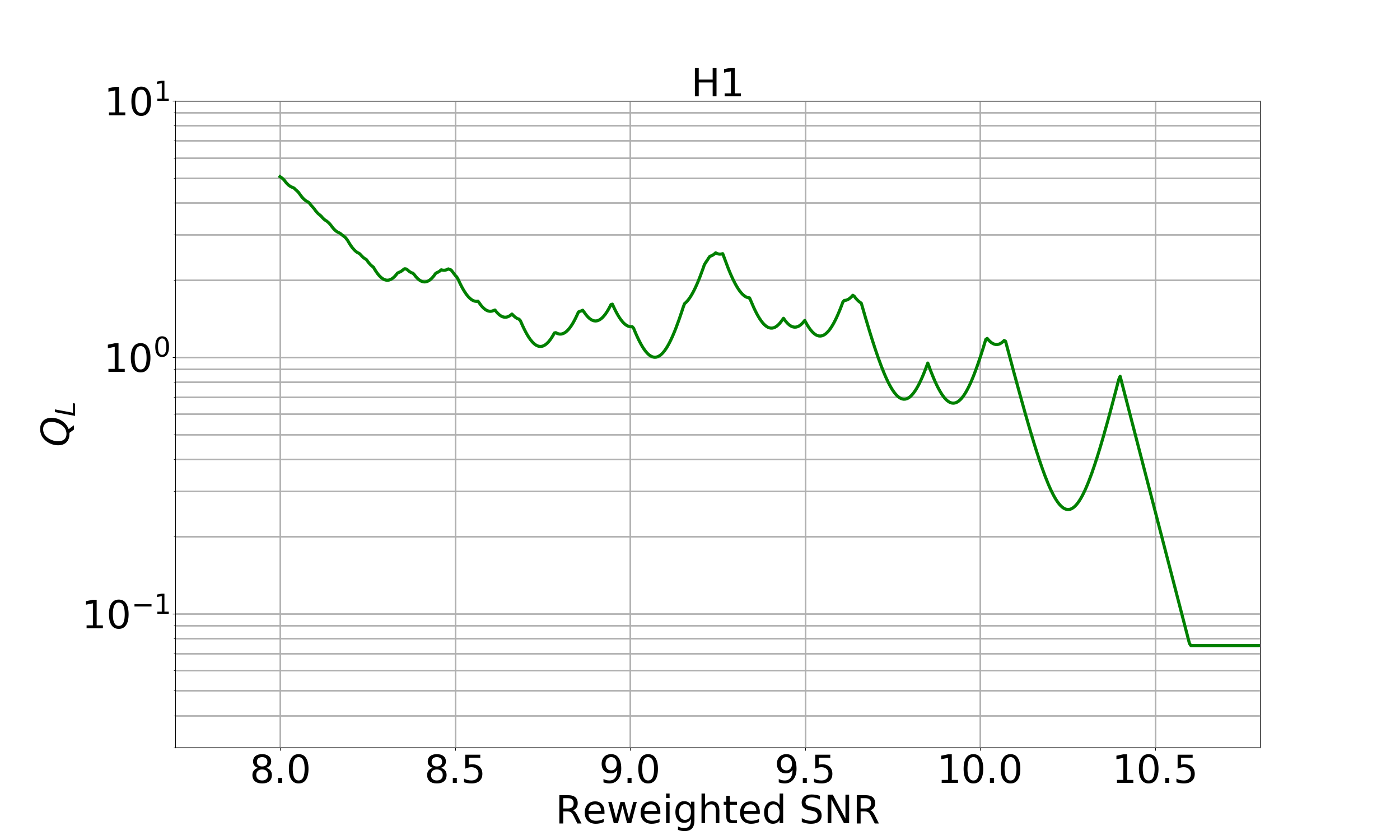}
   \end{minipage}\hfill
   \begin{minipage}{0.55\textwidth}
     \centering
     \includegraphics[width=\linewidth]{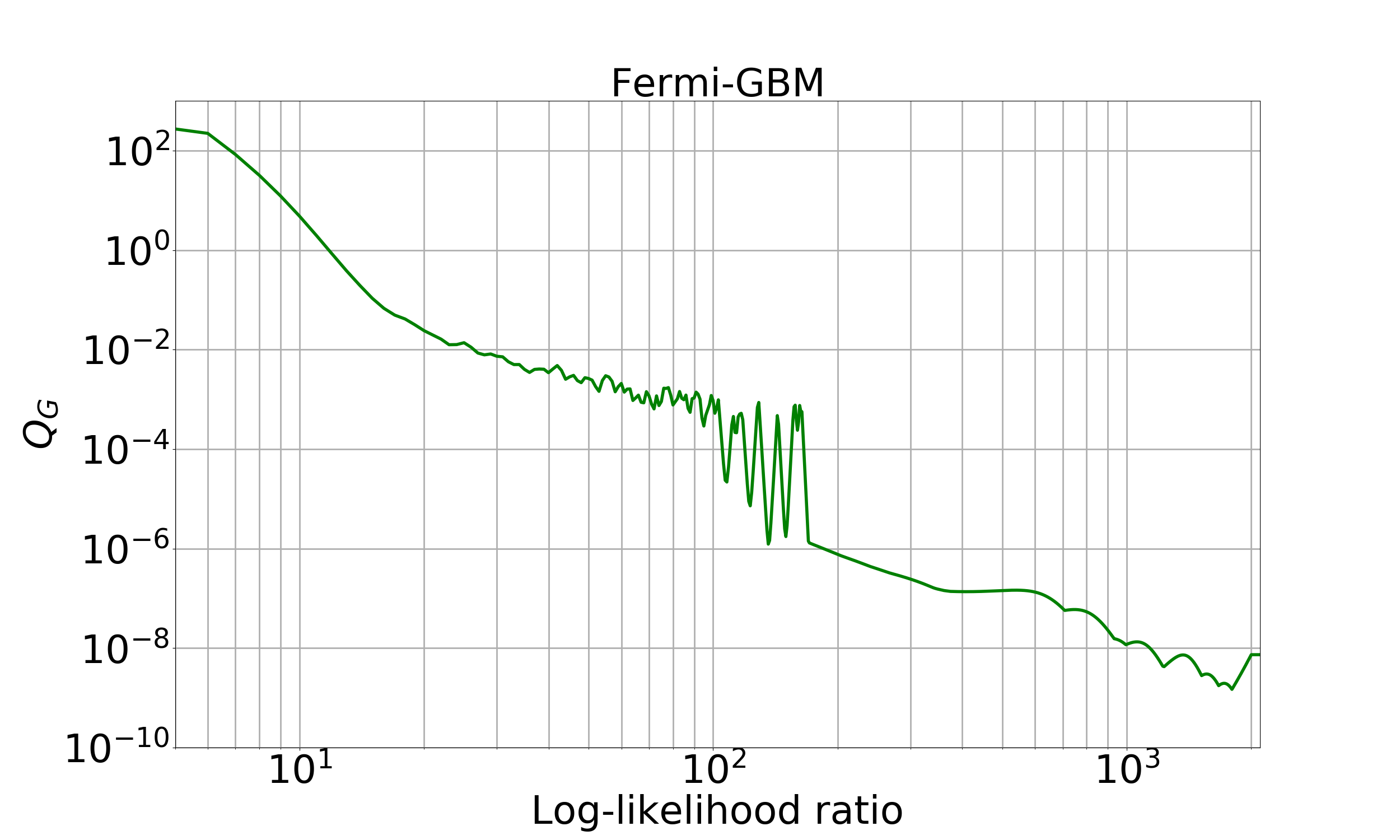}
   \end{minipage}
    \begin{minipage}{0.55\textwidth}
     \caption{On top O2 LIGO Bayes factor $Q_L$ for L1 (top left) and H1 (top right). On bottom Fermi-GBM Bayes factor $Q_G$.}
  \label{Fig:Bayes_factor}
   \end{minipage}
\end{figure}

Once the four distributions $P(LLR|\textrm{noise})$, $P(LLR|\textrm{signal})$, $P(\hat\rho_{gw}|\textrm{noise})$ and $P(\hat\rho_{gw}|\textrm{signal})$ have been calculated, the computation of the Bayes factors $Q_G(LLR)$ and $Q_L(\hat\rho_{gw})$ can be performed. The variation of the Bayes factors with the candidate parameters is shown in Figure~\ref{Fig:Bayes_factor}.
One can see that the curves present ``spikes'' or rapid oscillations. These features are artifacts due to using KDEs with a constant bandwidth in regimes where the data points are very sparse. An analysis of the impact of these artifacts on the final results can be found in \ref{sec:alternative_bayes}, where we show that the final results are unlikely to be affected by this behavior.

The spatial overlap term $I_{\mathbf{\Omega}}$ is calculated like in \cite{Ashton_2018}. While the Targeted Search provides a skymap for the Fermi-GBM candidate, for the GW trigger we generate a Bayestar skymap. Bayestar is a Bayesian localization algorithm~\cite{PhysRevD.93.024013} which has the advantage of rapidly (a few seconds) producing a reliable skymap without exploring the intrinsic source parameters as do Markov Chain Monte Carlo based methods of parameter estimation~\cite{Veitch2015}. Another detail to emphasize is that the Bayestar skymaps for single interferometer triggers are not very informative, as they simply follow the directional response of the interferometer. For a single interferometer skymap, the 50\% credible region covers around 8000 square degrees, whereas the 90\% credible region occupies approximately 24000 square degrees. If one notes by $D_{L}$ and $D_{G}$ the data from LIGO and Fermi-GBM and by $\mathbf{\Omega}$ the sky location of the source, the expression of the skymap overlap term is
\begin{equation}
    I_{\mathbf{\Omega}} = \int \frac{P(\mathbf{\Omega}|D_{L})P(\mathbf{\Omega}|D_{G})}{P(\mathbf{\Omega})}d\mathbf{\Omega}.
\end{equation}
We assume a uniform prior $P(\mathbf{\Omega}) = 1/(4\pi)$. It is worth mentioning that the Earth is already excluded in $P(\mathbf{\Omega}|D_{G})$. Note that if one of the data sets is poorly informative with respect to the sky location, i.e.~$P(\mathbf{\Omega}|D_{L|G}) \approx P(\mathbf{\Omega})$ for all $\mathbf{\Omega}$, then $I_{\mathbf{\Omega}} \approx 1$ regardless of the precision of the other sky localization.

\begin{figure}[!htb]
    \centering
    \includegraphics[scale=0.25]{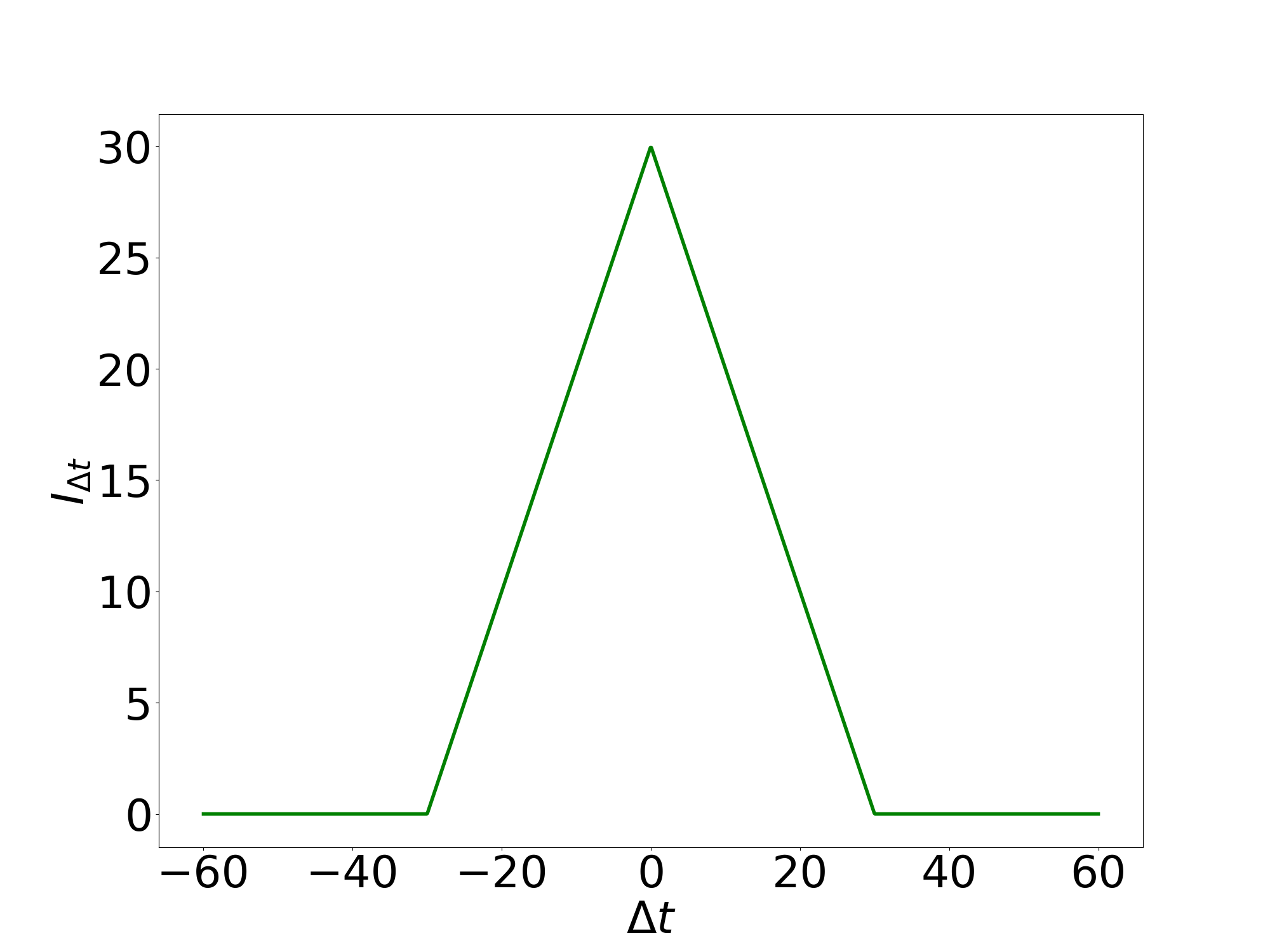}
    \caption{Time overlap term $I_{\Delta t}$ as a function of the time offset $\Delta t$.}
    \label{Fig:triangular}
\end{figure}

The time offset term $I_{\Delta t}$ accounts for how probable it is for a pair formed by a GW trigger and a Fermi-GBM trigger to be separated by a certain amount of time $\Delta t = t_{EM} - t_{GW}$, where $t_{GW}$ represents the estimated merger time of the GW candidate and $t_{EM}$ is the central time of the GBM trigger with the maximum LLR. We assume that the GWs and the EM waves travel at the same speed~\cite{abbott2017gravitational}, but there is not complete knowledge about the intrinsic time offset at the source. For this study, our choice is a search for which the offset term has a triangular shape (Figure~\ref{Fig:triangular}) centered on 0, i.e.
\begin{equation}
    I_{\Delta t} = \left\{ \begin{array}{ll}
        30 - \left| \frac{\Delta t}{1\ \mathrm{s}} \right| & \textrm{if}\ |\Delta t| < 30\ \textrm{s} \\
        0 & \textrm{otherwise.}
    \end{array} \right.
    \label{eq:timeterm}
\end{equation}
Other continuous and simple functional forms with similar support over $\Delta t$, symmetric around $\Delta t = 0$ and favoring small $\Delta t$ values would provide equally reasonable choices. Under these constraints, we do not expect our results to dramatically depend on the precise shape of the function.

\section{Calculation of a FAR}
\begin{figure}[!htb]
   \begin{minipage}{0.55\textwidth}
     \centering
     \includegraphics[width=\linewidth]{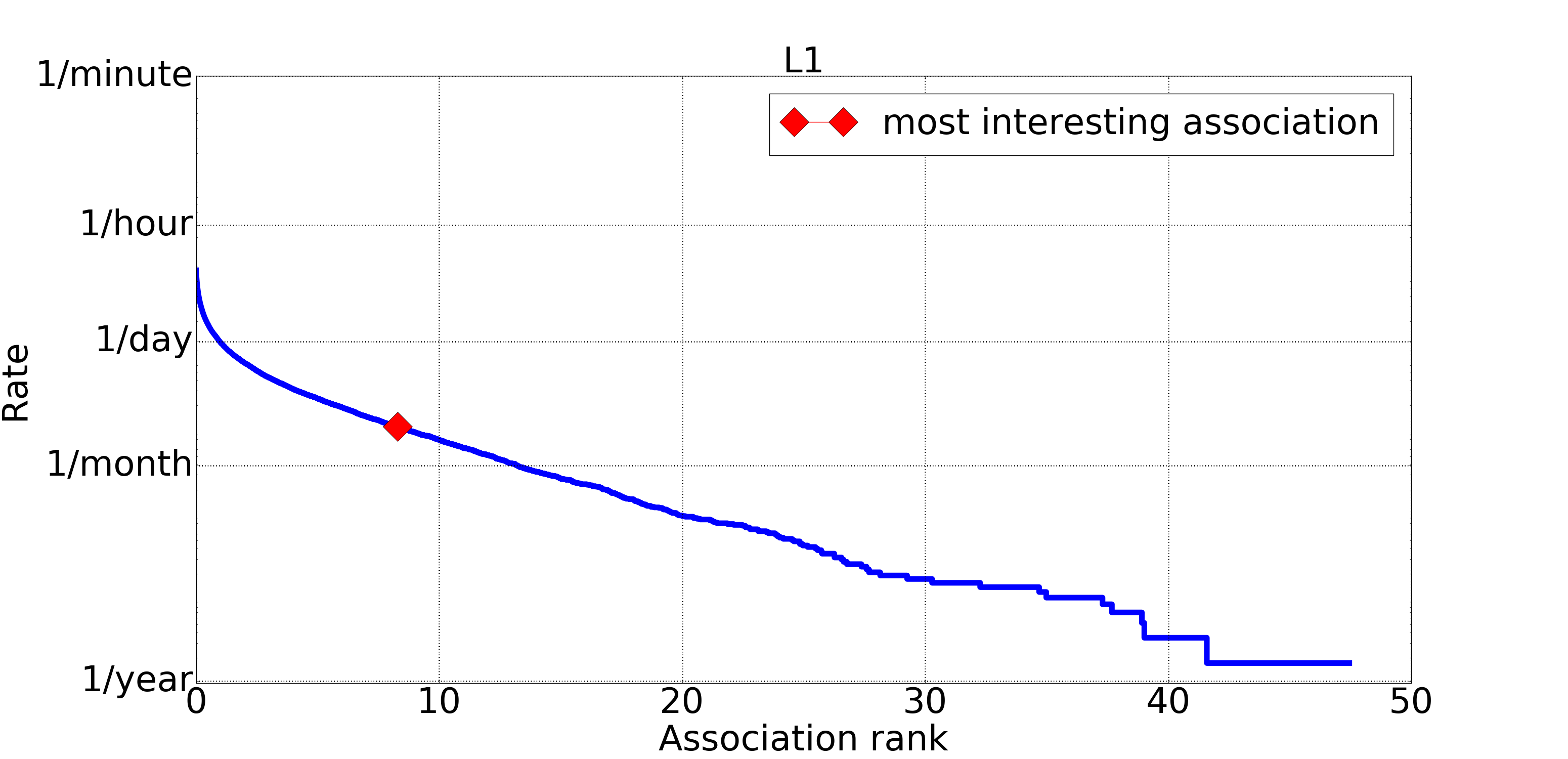}
   \end{minipage}\hfill
   \begin{minipage}{0.55\textwidth}
     \centering
     \includegraphics[width=\linewidth]{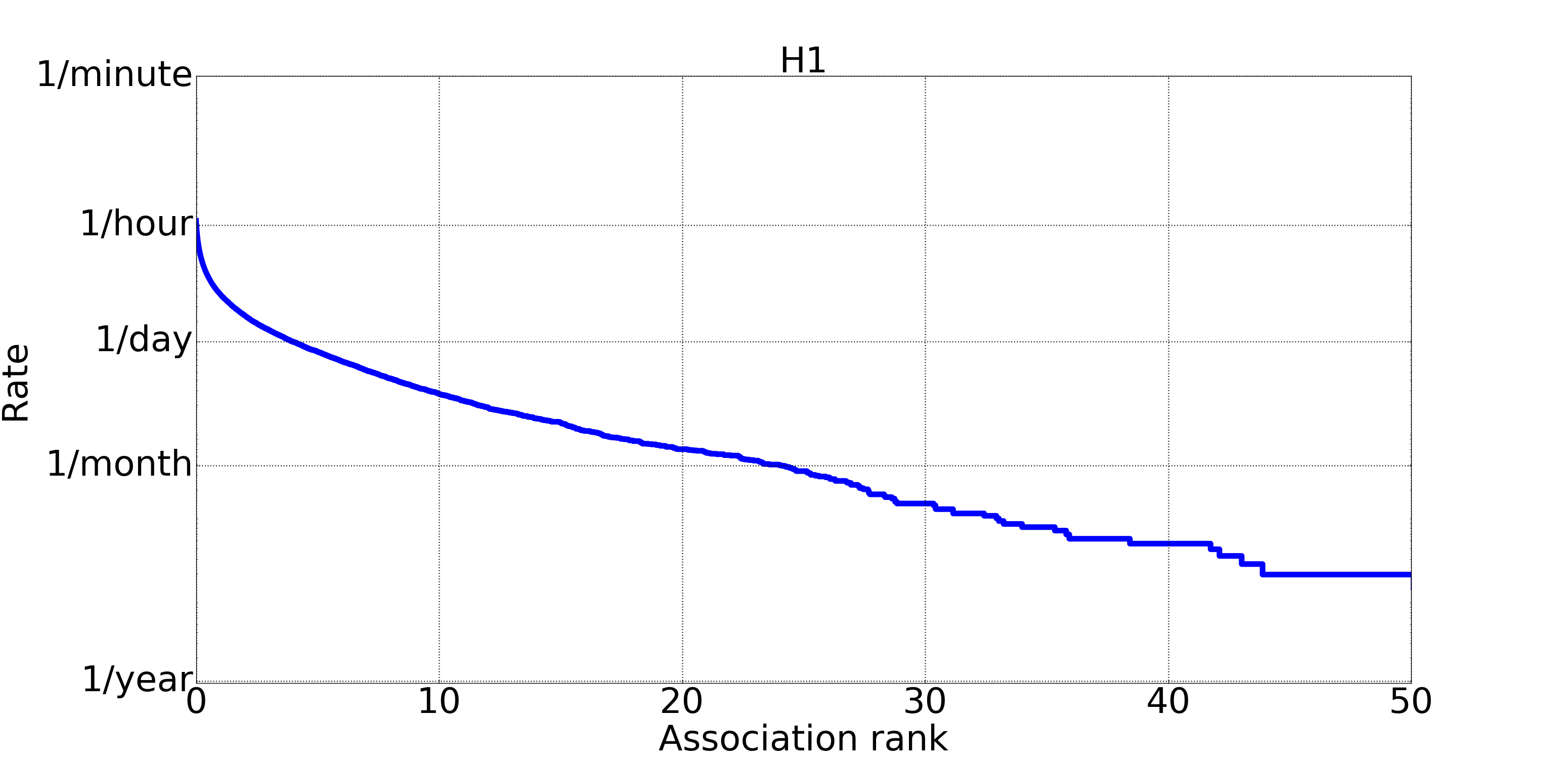}
   \end{minipage}
   \caption{$\Lambda$ distribution for pairs of time-shifted triggers between Fermi-GBM and LIGO-Livingston (left) or LIGO-Hanford (right). Within the left (L1) figure, the red diamond is drawn at the $\Lambda$ of the most interesting un-shifted event from the search; the vertical coordinate then indicates its FAR.}
   \label{Fig:FAR}
\end{figure}

Via an empirical estimation of its background distribution, $\Lambda$ is converted to a FAR, a quantity expressing how often two unrelated events (either due to signals from different sources, or noise) lead to a particular value of $\Lambda$ or a higher value. Methods to calculate FARs are ubiquitous in LIGO-Virgo data analysis, and are commonly based on time slides~\cite{Was:2009vh}. Here we start with a set of trigger candidates in both LIGO and Fermi-GBM data. The same set of GW triggers is used to generate both the foreground and the background. In the case of GBM triggers, the situation is different. For the GBM triggers used in the calculation of the FAR, we run the Targeted Search on consecutive 60 s time windows with the same configuration used to produce the foreground triggers. The background interval covers 23 days centered around GPS time 1180561923, the time of the most interesting candidate from our search (discussed later, see Section 5). Then we time-shift the resulting GBM triggers by a nonzero integer multiple of 50 s and we calculate the association ranking statistic again using the GW triggers and the time-shifted GBM triggers. We assume a $\pm 50$ s offset to be an unphysical time delay between a CBC and any possible GRB emission resulting from it, which is consistent with the maximum time offset considered in Eq (\ref{eq:timeterm}). We repeat this process multiple times, each with a different nonzero integer multiple of 50 s, and accumulate the background distribution of $\Lambda$ values, shown in Figure~\ref{Fig:FAR}, which provides a mapping between $\Lambda$ and FAR normalized by the total coincident GW-GBM live time resulting from the time shifts.

It is worth mentioning that this method of calculation of a FAR is different from just taking the distribution of foregrounds. In particular, the FAR of the loudest event is not simply the inverse of the observation time.

\section{Analysis of O1 and O2 data}
\label{sec:analysis}
For O1 and O2 we analyzed Fermi-GBM counterparts to all LIGO single interferometer PyCBC triggers having an $\hat\rho_{gw}$ higher than 8. That accounts for 1621 (1126 for O2, 495 for O1) such triggers. 

A first selection consists in considering the 80 candidates having the lowest FAR. For each of these triggers, LIGO detector characterization methods were applied. This qualitative analysis was performed by means of Omicron Scans and Used Percentage Vetoes~\cite{abbott2018effects, Abbott_2016, Isogai_2010}. The presence of known instrumental glitches, blip glitches~\cite{Abbott_2016,cabero2019blip}, stationary noise or scattered light represented a reason for rejection of 64 candidates. Twelve other candidates were also ignored because parameter estimation~\cite{Veitch2015} either returned a low ($<$ 5) log Bayes factor (little evidence for signal hypotheses), or showed evidence of bimodality in the posterior of different CBC parameters. Finally, noteworthy poor background fits in the low-energy channels of the GBM detectors represented the reason for the rejection of 3 other candidates.

\begin{figure}[!htb]
    \centering
    \includegraphics[scale=0.5]{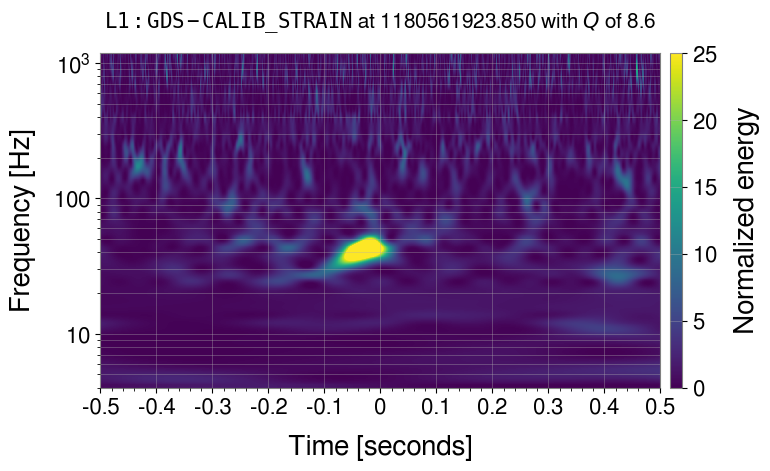}
    \caption{Spectrogram of LIGO-Livingston data around June 03, 2017 21:51:45 UTC, the time of the one remaining GW candidate from the single interferometer search.}
    \label{Fig:Omicron}
\end{figure}

At the end of the analysis described above there remains one mildly interesting association. This potential binary black hole merger signal was observed during O2 when only the Livingston interferometer was operating in science observing mode (Figure \ref{Fig:Omicron}).

\begin{figure}[!htb]
    \centering
    \includegraphics[scale=0.5]{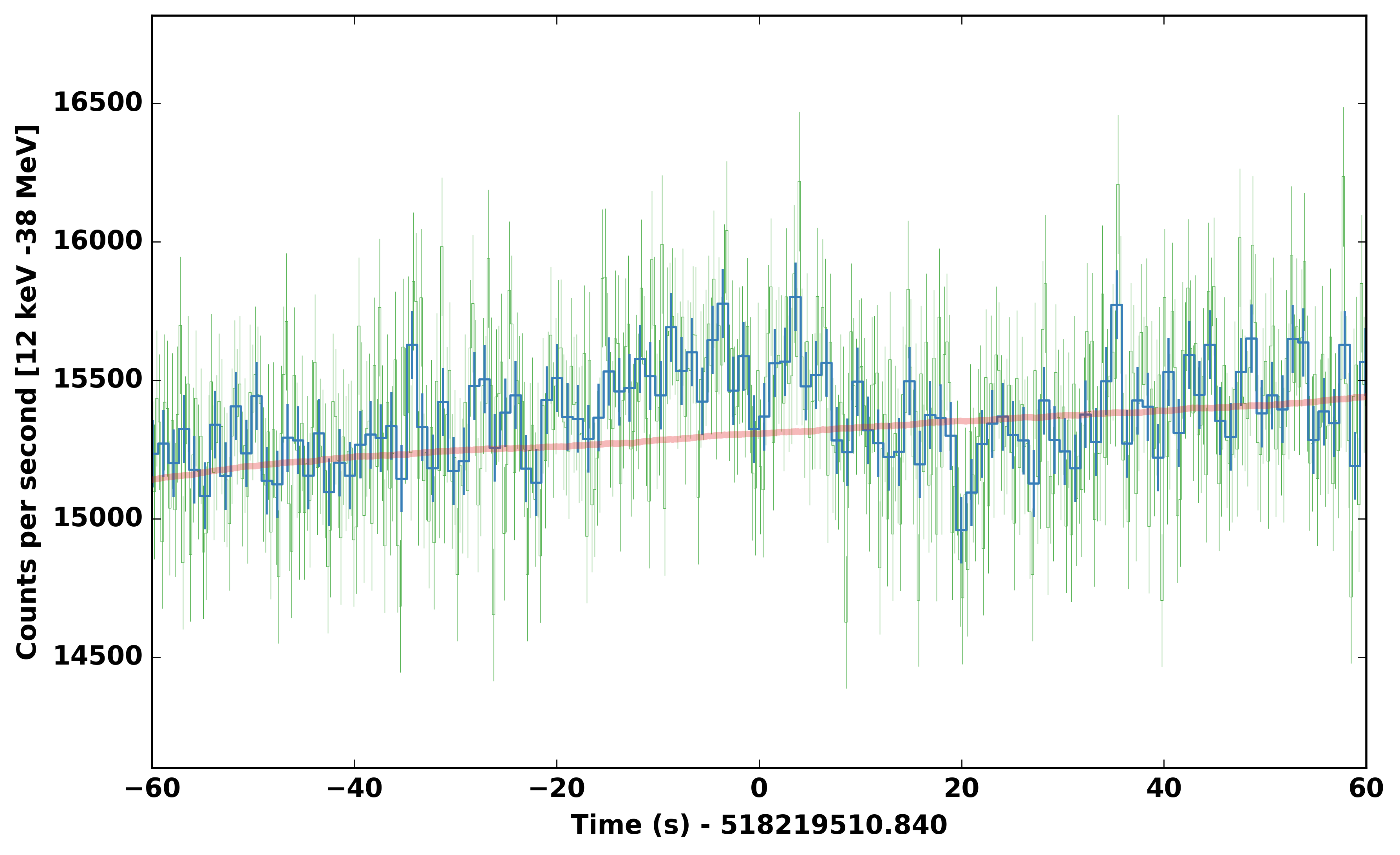}
    \caption{GBM lightcurve for the Targeted Search counterpart to the LIGO trigger on June 03, 2017 21:51:45 UTC. The lightcurve is summed over all 12 NaI detectors and energies between 12 keV and 38 MeV.}
    \label{Fig:lightcurve}
\end{figure}

PyCBC produced a trigger with $\hat\rho_{gw} = 9.04$. The duration of the signal is very short, therefore if it were a binary merger, it would have to have a total mass higher (more than 200 solar masses, as determined by~\cite{Veitch2015}) than any reported so far. The results from parameter estimation using LALInference~\cite{Veitch2015} provide a log Bayes Factor (signal to Gaussian noise) of 12.3. The Targeted Search detects a corresponding subthreshold candidate with $LLR = 30.63$.  The lightcurve, summed over all detectors, of the GBM candidate is shown in Figure \ref{Fig:lightcurve}. 

\begin{figure}[!htb]
   \begin{minipage}{0.55\textwidth}
     \centering
     \includegraphics[width=\linewidth]{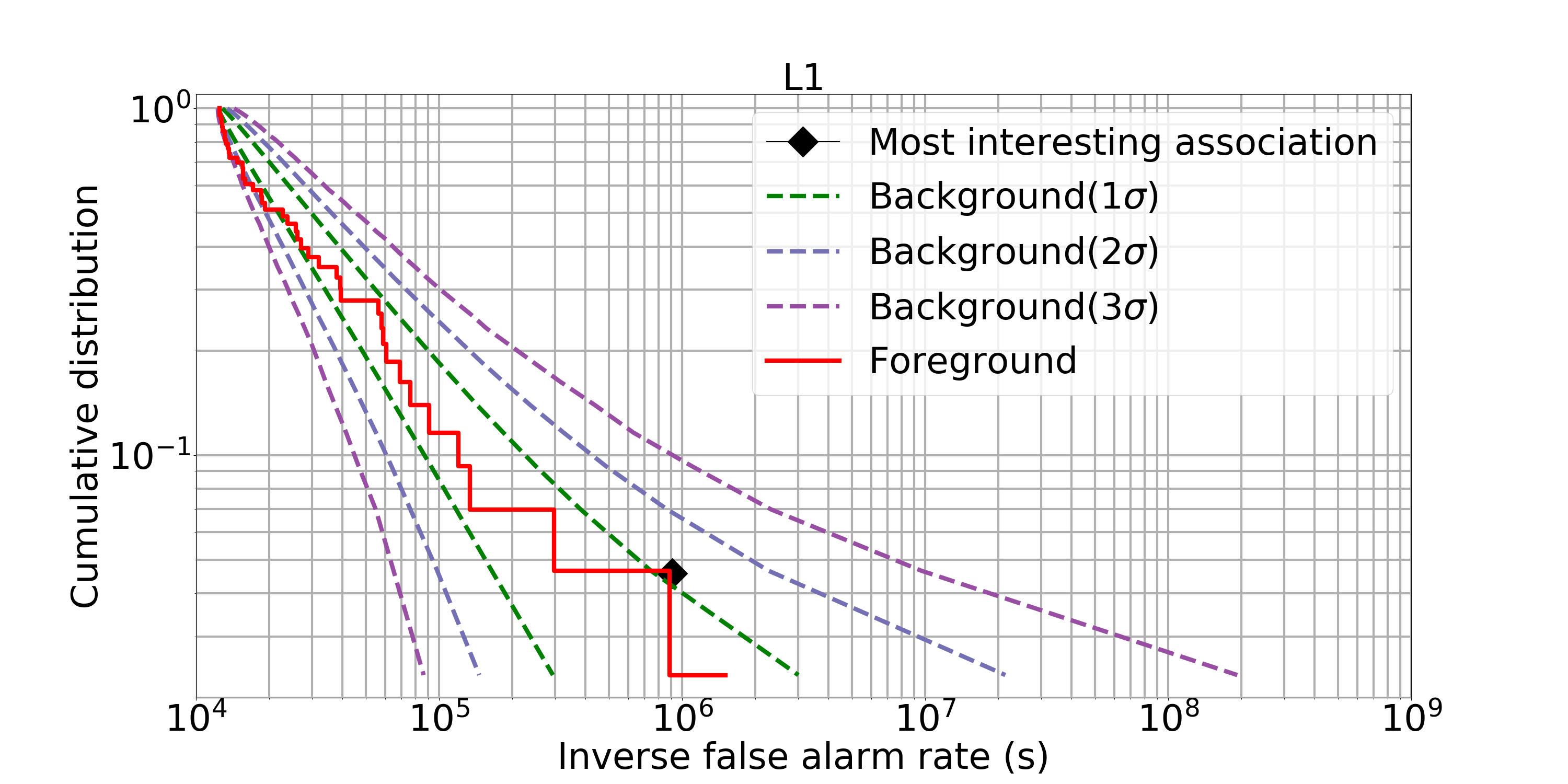}
   \end{minipage}\hfill
   \begin{minipage}{0.55\textwidth}
     \centering
     \includegraphics[width=\linewidth]{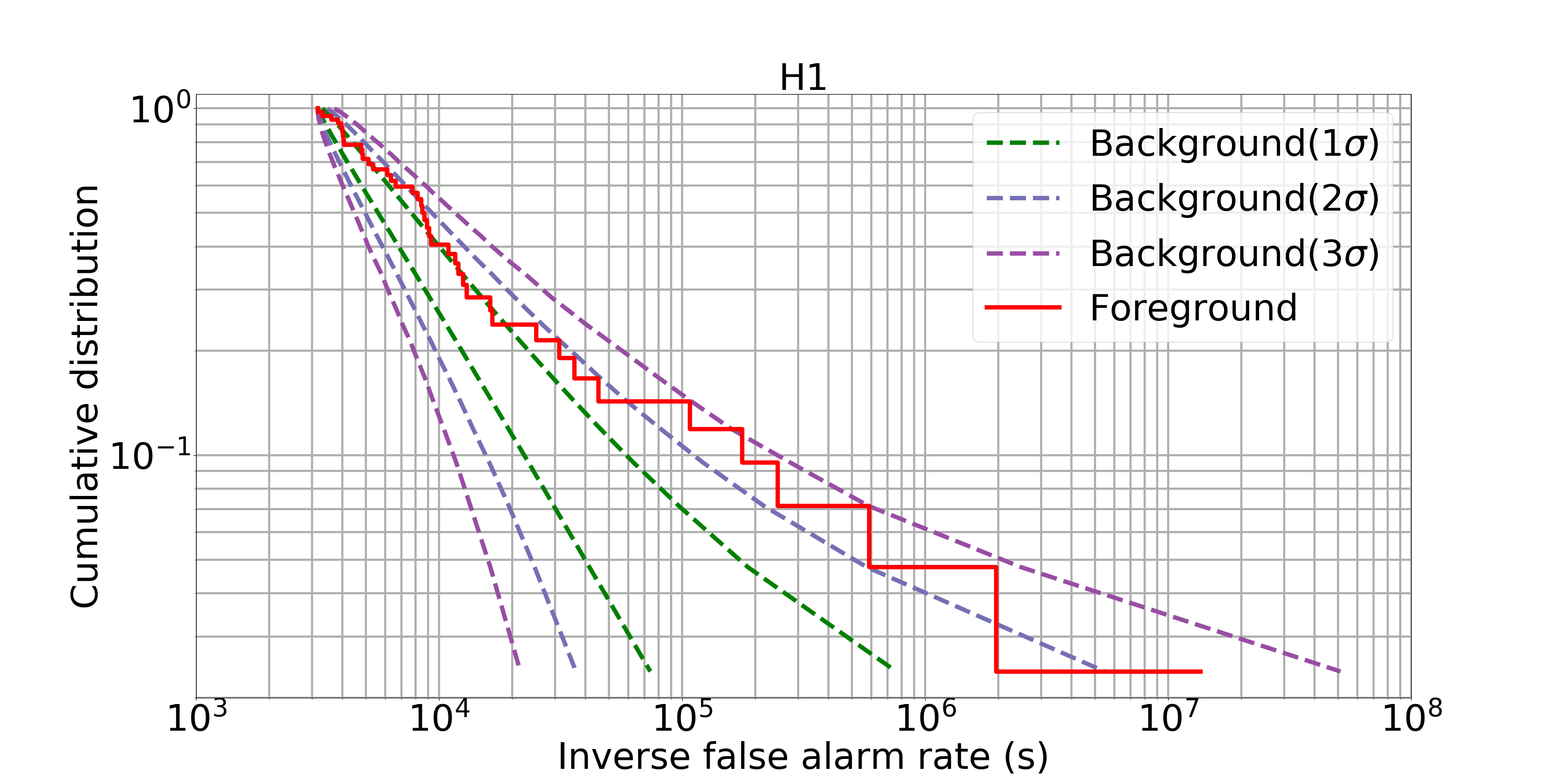}
   \end{minipage}
   \caption{Cumulative distribution function versus inverse false alarm rate (IFAR) for backgrounds assigned by uncertainties (in dashed lines) and foregrounds (in solid line). The results for L1 are at left and for H1 at right. The foregrounds represent associations between Fermi-GBM candidates and LIGO triggers with no time shift.  In the case of L1,  the black diamond represents the IFAR of our most interesting association.}
   \label{Fig:culm}
\end{figure}

Investigation of the GBM candidate reveals a soft spectrum and a localization consistent with the galactic plane. The candidate is likely produced by Scorpius X-1 as a strong occultation step resulting from this Galactic X-ray source was observed close in time to the trigger. Calculating the FAR as described in Section 3, we find a FAR of $1.1 \times 10^{-6}$ Hz for this association, or about 1 per 10 days, which is not significant. The association ranking statistic and FAR for this event are illustrated in Figure~\ref{Fig:FAR}. Presented in Figure~\ref{Fig:culm} are the cumulative distributions (for LIGO-Hanford and LIGO-Livingston) for both the foreground (i.e., the events we analyzed) and background events. From the plots it is clear that either all PyCBC triggers were noise triggers, or perhaps some were astrophysical signals with no GRB emission.

We want to attract the attention of the readers to the differences between Figure~\ref{Fig:FAR} and Figure~\ref{Fig:culm}. While in Figure~\ref{Fig:FAR} we show the one-to-one correspondence between ranking statistic and FAR (calculation based only on background, the foreground plays no role), in Figure~\ref{Fig:culm} we compare the inverse false alarm rate (IFAR) distribution of the background with the IFAR distribution of the foreground.
\section{Conclusion}
In this paper we have presented a method to follow up LIGO single interferometer GW triggers with data from Fermi-GBM. For each GW trigger we found the most significant GBM counterpart within a $\pm 30$ s window. Then each GW/Fermi-GBM trigger pair was analyzed by the method described above. The main part of the analysis is a statistical study in which each pair is assigned an association ranking statistic based on the significance of each candidate, the skymaps' overlap, and their separation in time. The objective of this quantitative analysis is the calculation of a FAR distribution. But the most statistically significant pairs were also submitted to a qualitative analysis where we looked at the LIGO data quality and indications of non-cosmological $\gamma$-ray sources. The method described in this paper was used to search for coincident GW and $\gamma$-ray events by Fermi GBM and LIGO-Virgo over the O1 and O2 observing runs~\cite{Hamburg:2020dtg}.

For the analysis of the O1 and O2 PyCBC single interferometer LIGO triggers there remained one event of interest, although not statistically significant. Similar search methods will be applied during Advanced LIGO's and Advanced Virgo's third observing run, O3, which started in April 2019. For future searches we have the intention to improve our statistical method. One way to do that would be to find new derivations for the LIGO and GBM Bayes factors, for example taking into account the GW signal morphology in the time-frequency plane and the proximity of the GBM sky localization to the Sun and/or galactic plane. Distance/energy budget estimates could also in principle be incorporated into the ranking statistic. We are also considering improvements to the calculation of the LIGO and GBM Bayes factors by means of KDEs, for instance by using adaptive-bandwidth KDE and by formulating better models which can smoothly extend the distributions where only few triggers are available.

\ack
We thank the anonymous referees for valuable suggestions about future improvements and comments which improved the clarity of the presentation.

The work of JB and NC was supported by NSF grants PHY-1806990 and PHY-1505373. The work of PS was supported by NSF grant PHY-1710286. TD and EB were supported by an appointment to the NASA Postdoctoral Program at the Goddard Space Flight Center, administered by Universities Space Research Association under contract with NASA. The authors are grateful for computational resources provided by the LIGO Laboratory and supported by National Science Foundation Grants PHY-0757058 and PHY-0823459.

This paper has been assigned LIGO Document Number P1900296.

\appendix
\section{Choice of lower limits on $LLR$ and $\hat\rho_{gw}$}
\label{sec:lower_limits}
\renewcommand{\thefigure}{A\arabic{figure}}
\setcounter{figure}{0}
Restricting the analysis to GW triggers having $\hat{\rho}_{gw} \geq 8$ and Fermi-GBM triggers with $LLR \geq 5$ is motivated by practical considerations.

\begin{figure}[!htb]
   \begin{minipage}{0.55\textwidth}
     \centering
     \includegraphics[width=\linewidth]{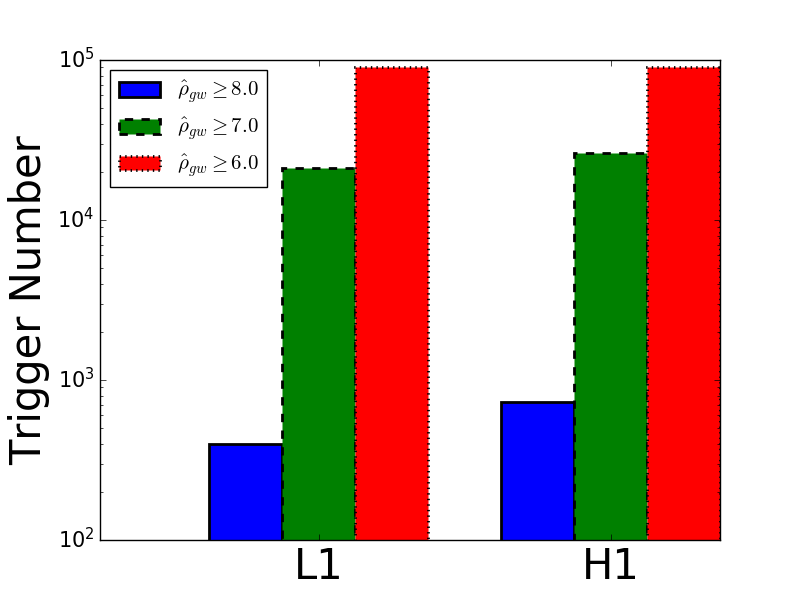}
   \end{minipage}\hfill
   \begin{minipage}{0.55\textwidth}
     \centering
     \includegraphics[width=\linewidth]{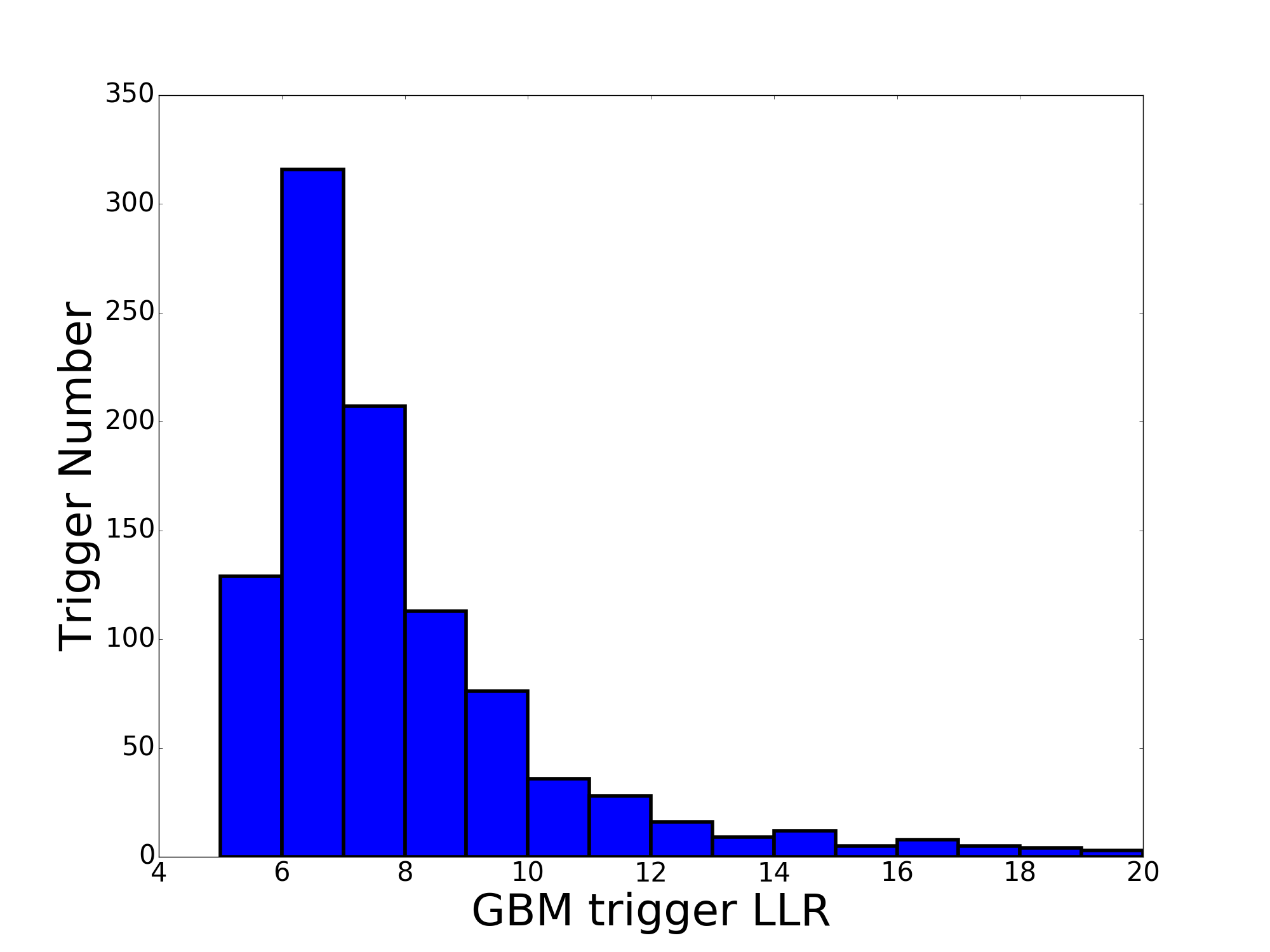}
   \end{minipage}
   \caption{On the left, the number of GW triggers having a reweighted SNR higher than 8 (in blue), higher than 7 (in green) and respectively higher than 6 (in red). On the right, the histogram of GBM triggers within $\pm 30$ s of GW triggers.}
   \label{Fig:fig_limits}
\end{figure}

Concerning the GW candidates, taking into consideration weaker triggers would significantly increase the computational cost of the method. In fact, in Figure~\ref{Fig:fig_limits} one can see that the number of PyCBC triggers with $\hat{\rho}_{gw} \geq 7$ is more than 10 times the number of triggers already considered. In this study we hence settled on the choice of 8, which is already deep into the distribution arising from the detector's quasistationary noise. It is possible that code optimizations might allow us to lower the threshold and consider a much larger number of weak triggers in the future.

Concerning the Fermi-GBM triggers, the choice of a $LLR$ threshold of $5$ was initially made based on similar computational considerations, but as we will show, it has no practical effect. Recall that for our foreground analysis we consider only the Targeted Search trigger having the highest $LLR$ among those found in the window $[t_{GW} - 30 \ \textrm{s}, t_{GW} + 30\ \textrm{s}]$, where $t_{GW}$ is the coalescence time of a GW trigger. Such triggers from the foreground analysis all have $LLR$ values already higher than $5$: the minimum $LLR$ is $5.73$. A histogram of $LLR$ is plotted in Figure~\ref{Fig:fig_limits} for completeness.

\section{Alternative LIGO Bayes factor}
\label{sec:alternative_bayes}
\renewcommand{\thefigure}{B\arabic{figure}}
\setcounter{figure}{0}
We have seen that the use of KDEs with a fixed bandwidth leads to large oscillations when the fitted data become very sparse. Here we investigate to what extent these fitting artifacts affect the final ranking of the associations.

In the case of GBM, these oscillations are noticeable in the GBM Bayes factor only in a relatively narrow range of high $LLR$ values, where the GBM Bayes factor consistently indicates a signal as opposed to noise. Although the Bayes factor rapidly oscillates by a couple orders of magnitude, it never approaches unity. Given that $\Lambda$ depends very weakly on $Q_G$ when $Q_G$ is much smaller than 1, we argue that this artifact has very little effect on the final ranking.

\begin{figure}[!htb]
   \begin{minipage}{0.55\textwidth}
     \centering
     \includegraphics[width=\linewidth]{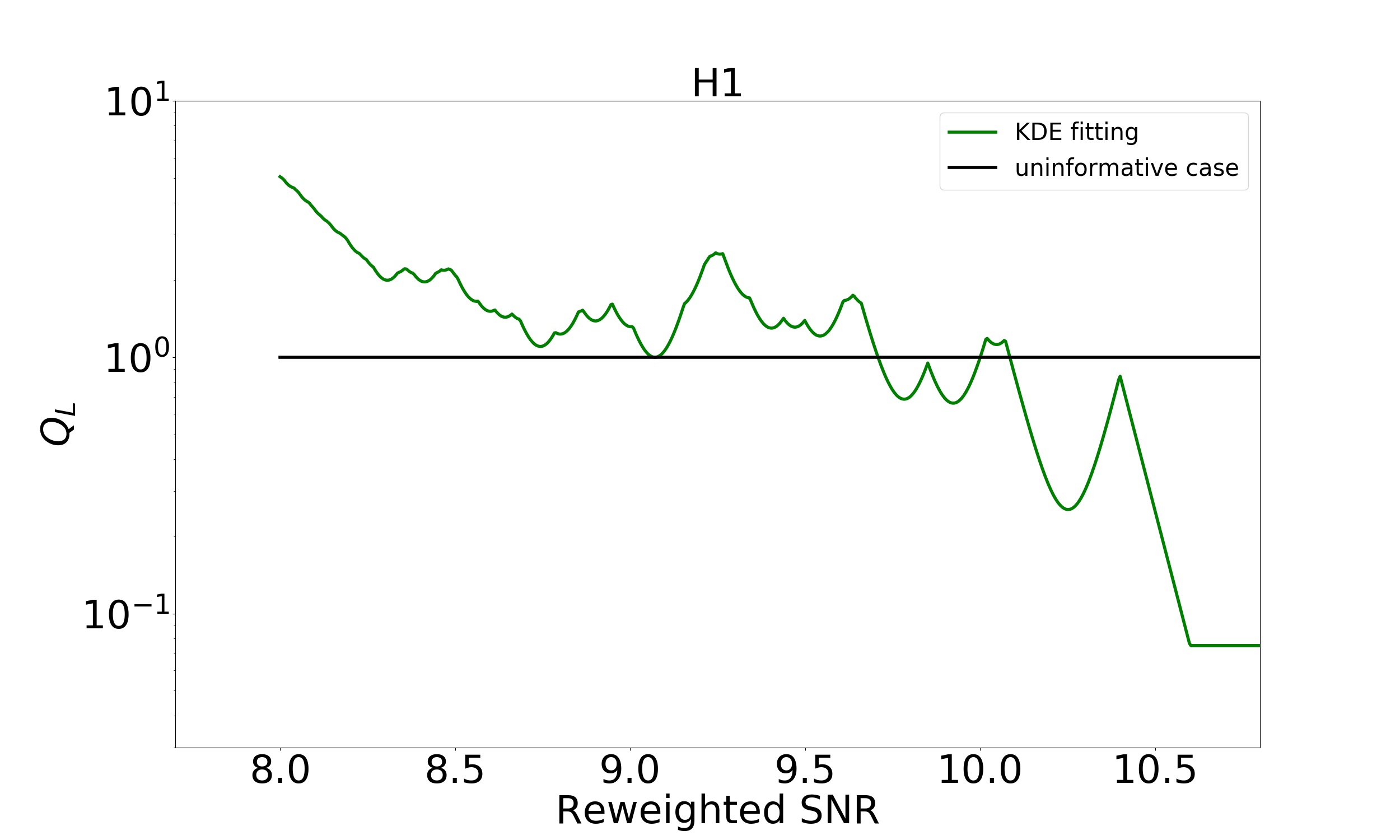}
   \end{minipage}\hfill
   \begin{minipage}{0.55\textwidth}
     \centering
     \includegraphics[width=\linewidth]{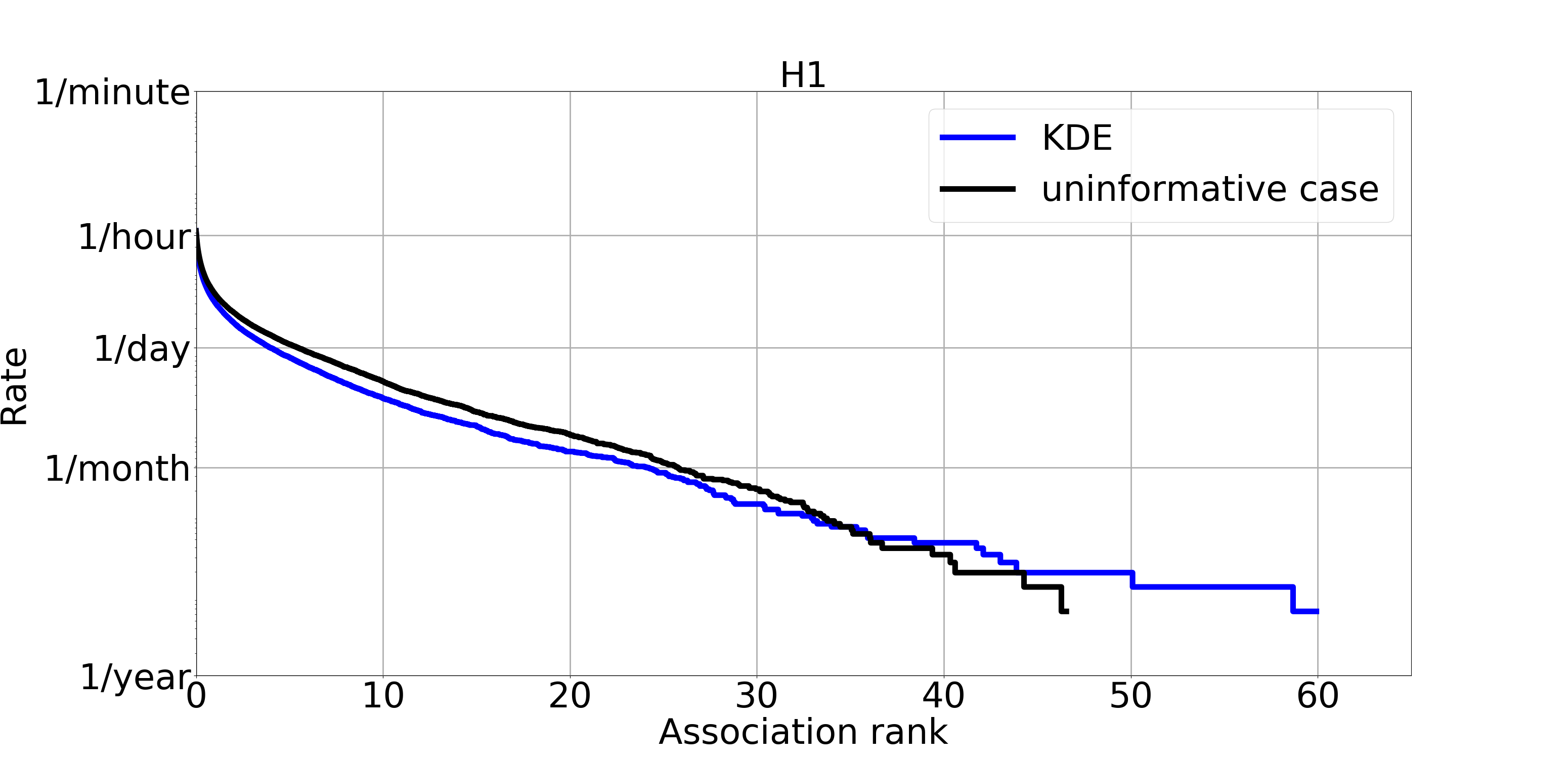}
   \end{minipage}
   \caption{On the left, the GW Bayes factor as a function of the reweighted SNR. In green the Bayes factor obtained by means of KDE (used in the analysis) and in black an uninformative Bayes factor, $Q_L(\hat\rho_{gw}) = 1$, for every $\hat\rho_{gw}$. On the right, the corresponding background FAR distributions. All these results correspond to H1.}
   \label{Fig:other_Bayes}
\end{figure}

On the other hand, in the case of GW triggers, the KDE oscillations are present in the regime of transition from weak to strong signals. One can also see in Figure~\ref{Fig:Bayes_factor}, however, that for the triggers we considered $Q_L(\hat\rho_{gw})$ has a relatively small dynamic range (less than two orders of magnitude) essentially centered around 1. The consequence of this behaviour is a ranking statistic $\Lambda$ that does not depend strongly on $\hat\rho_{gw}$. In order to check this statement, we can assign a completely uninformative GW Bayes factor $Q_L(\hat\rho_{gw}) = 1$ to every GW trigger and recalculate the background distribution of $\Lambda$. The result is shown in Figure~\ref{Fig:other_Bayes}. We can see that making $Q_L$ completely uninformative leads to a very similar background rate.

\begin{figure}[!htb]
   \begin{minipage}{0.55\textwidth}
     \centering
     \includegraphics[width=\linewidth]{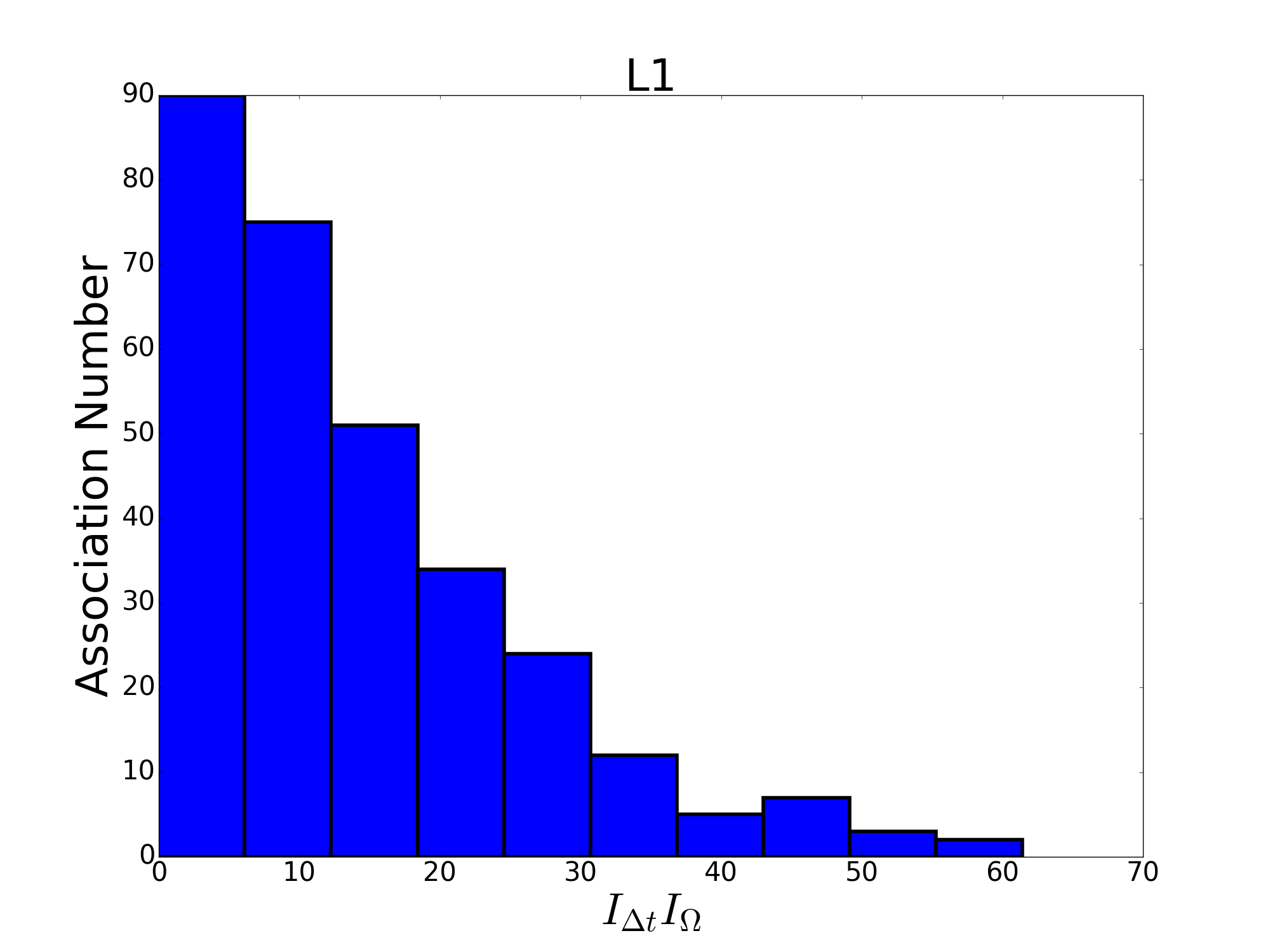}
   \end{minipage}\hfill
   \begin{minipage}{0.55\textwidth}
     \centering
     \includegraphics[width=\linewidth]{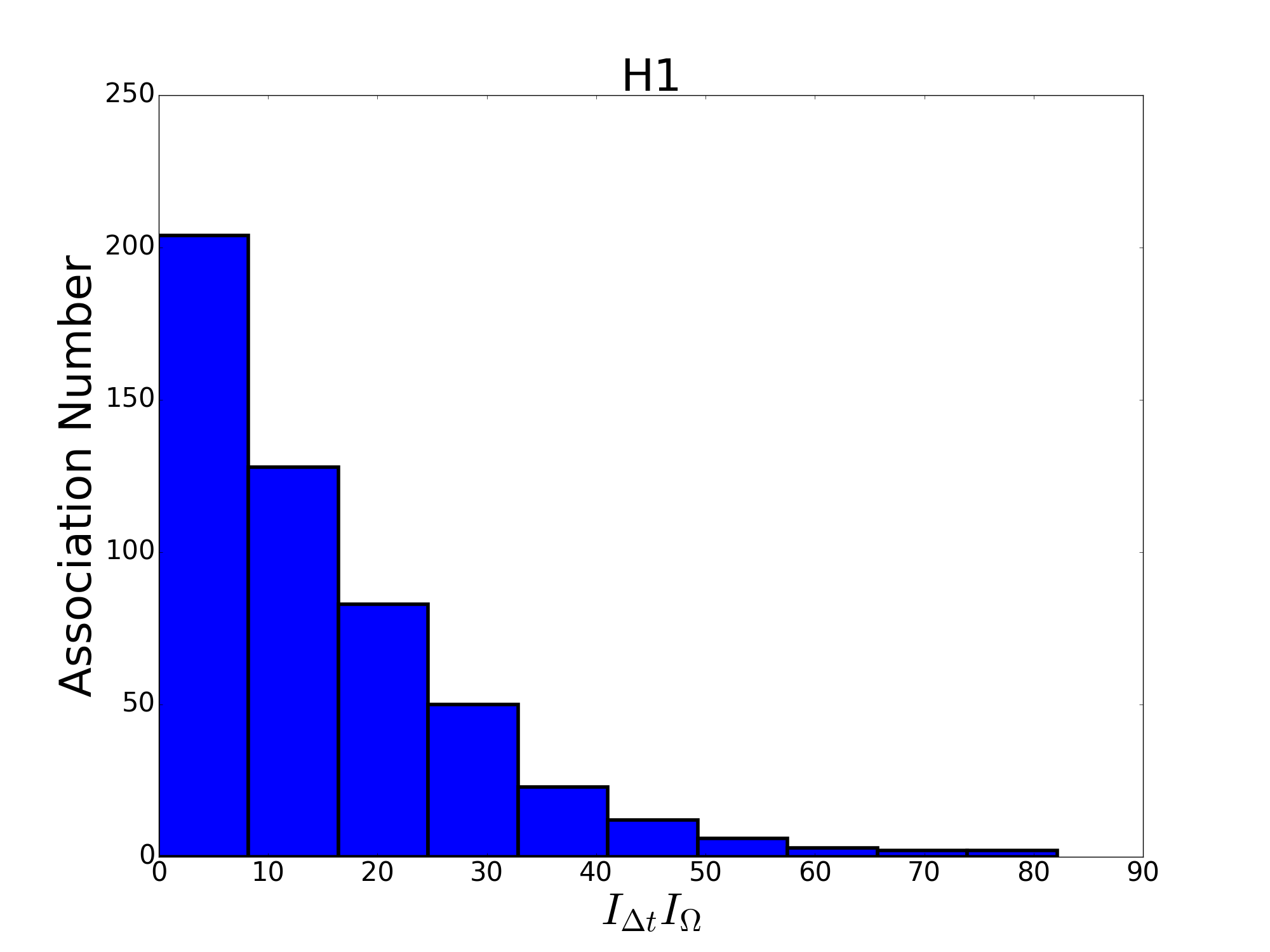}
   \end{minipage}
   \caption{The histogram of $I_{\mathbf{\Omega}}I_{\Delta t}$ for foreground. The L1 (H1) foreground is illustrated on left (right).}
   \label{Fig:numerator}
\end{figure}

A further question one can ask is if we might miss an interesting association due to the artifacts in the KDE fitting. From equation (12) we deduce that $\Lambda \leq I_{\mathbf{\Omega}}I_{\Delta t}$ for any $Q_L$ and $Q_G$. We can therefore inspect the $I_{\mathbf{\Omega}}I_{\Delta t}$ factors of the foreground candidates as an optimistic upper bound on their rank, under the assumption that their $Q_L$ values have been greatly overestimated due to the KDE artifacts. Figure~\ref{Fig:numerator} shows the distributions of $I_{\mathbf{\Omega}}I_{\Delta t}$ for our foreground associations.
There are just a few associations for which $I_{\mathbf{\Omega}}I_{\Delta t}$ is higher than 50, and they can easily be eliminated by means of LIGO and Fermi-GBM data quality arguments, such as described in Section~\ref{sec:analysis}. Lower values, given the background distributions in Figure~\ref{Fig:other_Bayes}, would all be associated to non-interesting FARs even if they had very small $Q_L$ values.

In conclusion, the quality of the distribution fits presented in Section~\ref{sec:association ranking} does not appear to be a limiting factor of this analysis. It could, however, become more problematic for the analysis of future data, and we are therefore investigating possible improvements.

\section*{References}
\bibliographystyle{iopart-num}
\bibliography{master}

\end{document}